\documentclass[twocolumn,prc,showpacs,preprintnumbers,superscriptaddress]{revtex4-1}

\usepackage{graphicx}% Include figure files
\usepackage{amsmath} %for text inside the equation,
\usepackage{amssymb}
\usepackage{amsfonts}
\usepackage{mathrsfs}
\usepackage{units}
\usepackage{multirow}

\usepackage{dcolumn}% Align table columns on decimal point
\usepackage{bm}% bold math
\usepackage{rotating} %Sideways tables

\renewcommand{\vec}[1]{\bm{#1}}
\newcommand{\disregard}[1]{}

\begin{document}

\title{Isospin-breaking corrections
to superallowed Fermi $\beta$-decay in isospin- and  angular-momentum-projected
nuclear Density Functional Theory}

\author{W. Satu{\l}a}
\affiliation{Institute of Theoretical Physics, Faculty of Physics, University of Warsaw, ul. Ho\.za
69, PL-00-681 Warsaw, Poland}

\author{J. Dobaczewski}
\affiliation{Institute of Theoretical Physics, Faculty of Physics, University of Warsaw, ul. Ho\.za
69, PL-00-681 Warsaw, Poland}
\affiliation{Department of Physics, P.O. Box 35 (YFL),
University of Jyv\"askyl\"a, FI-40014  Jyv\"askyl\"a, Finland}

 \author{W. Nazarewicz}
\affiliation{Department of Physics and
  Astronomy, University of Tennessee, Knoxville, Tennessee 37996, USA}
\affiliation{Physics Division, Oak Ridge National Laboratory, P.O. Box
  2008, Oak Ridge, Tennessee 37831, USA}
\affiliation{Institute of Theoretical Physics, Faculty of Physics, University of Warsaw, ul. Ho\.za
69, PL-00-681 Warsaw, Poland}

 \author{T.R. Werner}
\affiliation{Institute of Theoretical Physics, Faculty of Physics, University of Warsaw, ul. Ho\.za
69, PL-00-681 Warsaw, Poland}

\date{\today}

\begin{abstract}
\begin{description}
\item[Background]
The superallowed $\beta$-decay rates provide stringent constraints on physics beyond the Standard Model of
particle physics. To extract crucial information about the electroweak force,
small isospin-breaking corrections to the Fermi matrix element
of  superallowed transitions must be applied.
\item[Purpose]
We perform systematic calculations of isospin-breaking
corrections to  superallowed $\beta$-decays and estimate theoretical  uncertainties related to the basis truncation,
time-odd polarization effects related to the intrinsic symmetry of the underlying Slater
determinants, and to the functional parametrization.
\item[Methods]
We   use the self-consistent isospin- and  angular-momentum-projected  nuclear density functional theory  employing two density functionals derived from the density independent Skyrme interaction. Pairing correlations are ignored. Our  framework can
simultaneously describe  various effects that impact
matrix elements of the Fermi decay: symmetry breaking,
configuration mixing,  and long-range Coulomb polarization.
\item[Results]
The isospin-breaking corrections to
the $I=0^+,T=1\rightarrow I=0^+,T=1$ pure Fermi transitions are computed
for  nuclei from $A$=10 to $A$=98 and, for the first time, to
the Fermi branch of the $I,T=1/2 \rightarrow I,T=1/2$
transitions in mirror nuclei  from $A$=11 to $A$=49.
We carefully analyze various  model assumptions impacting
theoretical uncertainties of our calculations and provide theoretical error bars on our predictions.
\item[Conclusions]
The overall agreement with empirical isospin-breaking corrections is very satisfactory. Using computed isospin-breaking corrections we show that
the unitarity of the CKM matrix  is satisfied with a precision better than  0.1\%.

\end{description}
\end{abstract}

\maketitle

\section{Introduction}\label{intro}

By studying isotopes with enhanced sensitivity to fundamental symmetries, nuclear physicists can test various  aspects of the Standard Model in ways that are  complementary to other sciences.  For example, a possible explanation for the observed asymmetry between matter and anti-matter in the universe could be studied by searching for a permanent electric dipole moment larger than Standard Model predictions in heavy radioactive nuclei that have permanent octupole shapes. Likewise, the superallowed  $\beta$-decays of a handful of rare isotopes with similar numbers of protons and neutrons, in which both the parent and daughter nuclear states have zero angular momentum and positive parity, are the unique laboratory to study the strength of the weak force.

What makes these pure vector-current-mediated (Fermi) decays so useful for testing
the Standard Model is  the hypothesis of
the conserved vector current (CVC), that is, independence of the vector
current on the nuclear medium. The consequence of  the CVC hypothesis is that the product of the
statistical rate function $f$ and partial half-life $t$ for the superallowed
$I=0^+,T=1 \rightarrow I=0^+,T=1$ Fermi $\beta$-decay
should be nucleus independent and equal to:
\begin{equation}\label{ft}
      ft = \frac{K}{G_{\rm V}^2 |M_{\rm F}^{(\pm )}|^2 } = {\rm const}\, ,
\end{equation}
where $K/(\hbar c)^6 = 2\pi^3 \hbar \ln 2 /(m_{\rm e} c^2)^5 =
8120.2787(11)\times 10^{-10}$\,GeV$^{-4}$s  is
a universal constant; $G_{\rm V}$ stands for the vector coupling constant for
semi-leptonic weak interaction, and $M_{\rm F}^{(\pm )}$ is the
nuclear matrix element of the isospin rising or lowering
operator $\hat T_{\pm}$.

The relation (\ref{ft}) does not hold exactly and must be slightly amended
by introducing  a set of radiative corrections to the $ft$-values,
and a correction to the nuclear matrix element due to isospin-symmetry breaking:
\begin{equation}\label{MFa}
|M_{\rm F}^{(\pm )}|^2 = 2 (1- \delta_{\rm C} ),
\end{equation}
see Refs.~\cite{(Har05),*(Har05a),(Tow08),(Har09),(Tow10)}
and references cited therein. Since these corrections are small,  of the order of a percent,
they can be approximately factorized and arranged in the following way:
\begin{equation}\label{ftnew}
   {\cal F}t \equiv ft(1+\delta_{\rm R}^\prime)(1+\delta_{\rm NS} -\delta_{\rm C})
  = \frac{K}{2 G_{\rm V}^2 (1 + \Delta^{\rm V}_{\rm R})},
\end{equation}
with the left-hand side being nucleus independent. In Eq.~(\ref{ftnew}),
$\Delta^{\rm V}_{\rm R} = 2.361(38)$\% stands for the
nucleus-independent part of the radiative correction~\cite{(Mar06b)}, $\delta_{\rm R}^\prime$ is
a transition-dependent ($Z$-dependent) but nuclear-structure-independent part of
the radiative correction~\cite{(Mar06b),(Tow08)}, and $\delta_{\rm NS}$ denotes
the nuclear-structure-dependent part of the radiative
correction~\cite{(Tow94),(Tow08)}.

In spite of theoretical uncertainties in the evaluation of the radiative and
isospin-symmetry-breaking corrections, the superallowed $\beta$-decay
is the most precise source of experimental information for determining the vector
coupling constant $G_{\rm V}$, and provides us with a
stringent test of the CVC hypothesis. In turn, it is also the most precise
source of the matrix element $V_{\rm ud} = G_{\rm V}/G_{\rm \mu}$ of the Cabibbo-Kobayashi-Maskawa
(CKM) three-generation quark mixing
matrix~\cite{(Cab63),(Kob73),(Tow08),(Nak10)}. This is so because
the leptonic coupling constant, $G_{\mu}/(\hbar c)^3 = 1.16637(1)\times
10^{-5}$\, GeV${^{-2}}$, is well known from the muon decay~\cite{(Nak10)}.

The advantage of the superallowed $\beta$-decay strategy results from the fact that, within the CVC
hypothesis, $V_{\rm ud}$ can be extracted by averaging over several
 transitions in different nuclei. For precise tests of the Standard Model, only these transitions that have $ft$-values known with a relative
precision  better than  a fraction of a percent are acceptable.
Currently, 13 ``canonical" transitions spreading over a wide range of nuclei from
$A=10$ to $A=74$  meet this criterion (have $ft$-values measured with accuracy
of order of 0.3\%  or better) and are used to evaluate the values of $G_{\rm V}$ and
$V_{\rm ud}$~\cite{(Tow08)}.

In this work we concentrate on the isospin-breaking (ISB) corrections
$\delta_{\rm C}$, which were already computed by various authors,
using a  diverse set of nuclear
models \cite{(Dam69a),(Tow08),(Sag96a),(Lia09),(Aue09),(Sat11b),(Sat11c),(Sat11d),(Raf12)}.
The standard in this field has been set by  Towner and Hardy (HT)~\cite{(Tow08)} who used
the nuclear shell-model to account for the configuration mixing effect, and the mean-field (MF) approach
to account for a radial mismatch of proton and neutron
single-particle (s.p.) wave functions caused by the Coulomb polarization. In this study, which constitutes an extension of our earlier  work \cite{(Sat11c)}, we use the
isospin- and angular-momentum-projected  density functional theory (DFT).
This method  can account, in a rigorous quantum-mechanical  way, for spontaneous symmetry-breaking (SSB)
effects, configuration mixing, and long-range Coulomb polarization effects.

Our paper is organized as follows. The model is described in Sec.~\ref{model}.
The results of calculations for ISB corrections to the superallowed ${0^+  \rightarrow 0^+}$ Fermi transitions are summarized in Sec.~\ref{sec03}.
The ISB corrections to the Fermi matrix elements
in mirror-symmetric ${T=1/2}$ nuclei are discussed in Sec.~\ref{sec05}. Section \ref{sec06} studies a particular case of the Fermi decay  of $^{32}$Cl. Finally, the summary and perspectives are given in Sec.~\ref{sec07}.

\section{The model}\label{model}

The success of the self-consistent DFT approach
to mesoscopic systems~\cite{(Yan07)} in general, and specifically to
atomic nuclei~\cite{(Fra01),(Ben03),(Sat05)}, has
its roots in the SSB mechanism that incorporates essential short-range (pairing) and long-range (spatial) correlations
within a single deformed Slater determinant.
The deformed states provide a  basis for the symmetry-projected DFT approaches,
which aim at including beyond-mean-field correlations through the restoration
of  broken symmetries by means of projection techniques~\cite{(RS80)}.

\subsection{Isospin- and angular-momentum-projected DFT approach}
The building block of the isospin- and angular-momentum-projected DFT approach employed in this study
is the self-consistent deformed MF state $|\varphi \rangle$
that violates both the rotational and isospin symmetries.
While the rotational invariance is of fundamental nature
and is broken spontaneously, the isospin symmetry is
violated both spontaneously and explicitly by
the Coulomb interaction between protons.
The strategy is to restore the rotational invariance,
remove the spurious isospin mixing caused by the isospin SSB
effect, and retain only the physical isospin mixing
due to the electrostatic interaction~\cite{(Sat09a),(Sat10)}.  This is
achieved by a rediagonalization of the entire Hamiltonian,
consisting the isospin-invariant kinetic energy and Skyrme force and
the isospin-non-invariant Coulomb force,
in a basis that conserves both angular momentum and isospin.

To this end, we first find the self-consistent MF state
$|\varphi \rangle$ and then build a normalized angular-momentum-
and isospin-conserving basis $|\varphi ;\, IMK;\, TT_z\rangle$
by using the projection method:
\begin{equation}\label{ITbasis}
|\varphi ;\, IMK;\, TT_z\rangle =   \frac{1}{\sqrt{N_{\varphi;IMK;TT_z}}}
\hat P^T_{T_z,T_z} \hat P^I_{M,K} |\varphi \rangle ,
\end{equation}
where $\hat P^T_{T_z ,T_z}$ and $\hat P^I_{M,K}$ stand for the standard isospin
and angular-momentum projection operators:
\begin{eqnarray}
\hat P^T_{T_z, T_z} & = & \frac{2T+1}{2} \int_0^\pi
d^{T}_{T_z T_z}(\beta_T ) \hat{R}(\beta_T ) \sin\beta_T\, d\beta_T,   \\
\hat P^I_{M, K} & = & \frac{2I+1}{8\pi^2 } \int
 D^{I\, *}_{M K}(\Omega ) \hat{R}(\Omega ) \, d\Omega,
\end{eqnarray}
where, $\hat{R}(\beta_T )= e^{-i\beta_T \hat{T}_y}$
is the rotation operator
about the $y$-axis in the isospace, $d^{T}_{T_z T_z}(\beta_T )$ is
the Wigner function, and $T_z =(N-Z)/2$ is the third component
of the total isospin $\bm{T}$.  As usual, $\hat{R}(\Omega )= e^{-i\gamma \hat{J}_z}
e^{-i\beta \hat{J}_y} e^{-i\alpha \hat{J}_z}$ is the three-dimensional
rotation operator in space,  $\Omega = (\alpha, \beta, \gamma )$  are the
Euler angles, $D^{I}_{M K}(\Omega )$ is the Wigner function,
and $M$ and $K$ denote the angular-mo\-men\-tum components along the
laboratory and intrinsic $z$-axis, respectively \cite{(RS80),(Var88)}.
Note that unpaired MF
states $|\varphi \rangle$ conserve  the third isospin component
$T_z$; hence,  the one-dimensional isospin projection suffices.

The set of states (\ref{ITbasis}) is, in general, overcomplete because
the $K$ quantum number is not conserved. This difficulty is overcome
by selecting first the subset of linearly independent states
known  as {\it collective space} \cite{(RS80)}, which
is spanned, for each $I$ and $T$, by the so-called {\it natural states\/}
$|\varphi;\, IM;\, TT_z\rangle^{(i)}$ \cite{(Dob09d),(Zdu07a)}.
The entire Hamiltonian -- including
the ISB terms -- is rediagonalized in the collective space, and the resulting
eigenfunctions are:
\begin{equation}\label{KTmix}
|n; \,\varphi ; \,
IM; \, T_z\rangle =  \sum_{i,T\geq |T_z|}
   a^{(n;\varphi)}_{iIT} |\varphi;\, IM; TT_z\rangle^{(i)} ,
\end{equation}
where the index $n$ labels the eigenstates in ascending order
according to their energies. The amplitudes  $a^{(n;\varphi)}_{iIT}$
define the degree of isospin mixing through the so-called isospin-mixing
coefficients (or isospin impurities), determined for a given $n$th eigenstate as:
\begin{equation}
\label{truemix}
\alpha_{\rm C}^n = 1 - \sum_i |a^{(n;\varphi)}_{iIT}|^2,
\end{equation}
where the sum of norms
corresponds to the isospin $T$ dominating in the wave function $|n; \,\varphi ; \,
IM; \, T_z\rangle$.

One of the advantages of the projected DFT as compared to the
shell-model-based approaches~\cite{(Orm95a),(Tow08)} is that it allows
for a rigorous quantum-mechanical evaluation of the Fermi matrix element
using the bare isospin operators:
\begin{equation}
    \hat T_{\pm} = \frac{1}{2} \sum_{k=1}^A \left( \hat \tau^{(k)}_x \pm i \hat \tau^{(k)}_y \right)
    \equiv \mp \frac{1}{2} \hat T_{1\,\pm 1},
\end{equation}
where $\hat T_{1\,\pm 1}$ denotes the rank-one covariant
one-body spherical-tensor operators in the isospace, see the discussion in
Ref.~\cite{(Mil08), *(Mil09)}. Indeed,
noting that each $m$th eigenstate (\ref{KTmix}) can be uniquely decomposed
in terms of the original basis states (\ref{ITbasis}),
\begin{equation}\label{states}
|m; \varphi;\, IM;\, T_z\rangle = \sum_{K,T} \, f_{K T }^{(\varphi ;\,m,I)}
\hat P^T_{T_z, T_z} \, \hat P^I_{M, K} |\varphi \rangle ,
\end{equation}
with microscopically determined mixing coefficients $f_{KT}^{(\varphi ; \,m,I)}$,
the expression for the Fermi matrix element between the parent state $|m; \,\varphi ; \, IM; \, T_z\rangle$  and
daughter state $|n; \,\psi ; \, IM; \, T_z\pm 1 \rangle$ can be written as:
\begin{widetext}
\begin{eqnarray}\label{mate}
  \langle m; \,\varphi ; \, IM; \, T_z | \hat T_{\mp } |
   n; \,\psi ; \, IM; \, T_z\pm 1 \rangle = \pm \frac{1}{2}
   \sum_{TT'} \sum_{KK'} f_{KT}^{(\varphi ; \,m,I)\, *} \, f_{K'T'}^{(\psi ; \,n,I)}
   \langle \varphi | \hat P^T_{T_z, T_z}  \hat T_{1\,\mp 1} \hat P^{T'}_{T_z\pm 1, T_z\pm 1} \hat P^I_{K, K'} |\psi \rangle \nonumber \\
   = \pm \frac{2I+1}{16\pi^2} \sum_{TT'} \sum_{KK'} f_{KT}^{(\varphi ; \,m,I)\, *} \, f_{K'T'}^{(\psi ; \,n,I)}
   \int d\Omega \, D^{I\, *}_{KK'} (\Omega)
   \langle \varphi | \hat P^T_{T_z, T_z}  \hat T_{1\,\mp 1} \hat P^{T'}_{T_z\pm 1, T_z\pm 1} |{\tilde \psi} \rangle ,
   \end{eqnarray}
%\end{widetext}
where tilde indicates the Slater determinant rotated in space:
$|{\tilde \psi} \rangle = |\psi(\Omega )\rangle  = \hat R(\Omega ) | \psi \rangle$.
The matrix element appearing on the right-hand side of Eq.~(\ref{mate})
can be expressed through  the transition densities that are  basic building blocks
of the multi-reference DFT~\cite{(She00a),(Ang01),(Lac09),(Sat10)}.
Indeed, with the aid of the identity
\begin{eqnarray}
%\begin{equation}
 &\, & \hat P^{T}_{K, M}  \hat T_{\lambda \, \mu} \hat P^{T'}_{M',K'} =
% \nonumber \\ &\, &
  C^{T M}_{T' M'\, \lambda \mu} \sum_{\nu = -\lambda }^\lambda C^{T K}_{T' K-\nu \, \lambda \nu}
  \hat T_{\lambda \nu } \hat P^{T'}_{K-\nu, K'} ,
%\end{equation}
\end{eqnarray}
%\begin{eqnarray}
%  &\, & \hat P^T_{T_z, T_z}  \hat T_{1\,\mp 1} \hat P^{T'}_{T_z\pm 1, T_z\pm 1} = \nonumber \\
%  &\, & C^{TT_z}_{T'T_z\pm 1\, 1\mp 1} \sum_\mu C^{TT_z}_{T'T_z -\mu\, 1\mu} \hat T_{1\mu }
%  \hat P^{T'}_{T_z - \mu, T_z\pm 1} ,
%\end{eqnarray}
which results from the general transformation rule for spherical tensors under rotations
or isorotations,
\begin{equation}\label{trans}
\hat R(\Omega ) \hat T_{\lambda\mu} {\hat R(\Omega )}^\dagger = \sum_{\mu'} D^\lambda_{\mu' \mu} (\Omega )
\hat T_{\lambda \mu'} ,
\end{equation}
the matrix element entering  Eq.~(\ref{mate}) can be expressed as:
\begin{eqnarray}\label{mate1}
&\,& \langle \varphi | \hat P^T_{T_z, T_z}  \hat T_{1\,\mp 1} \hat P^{T'}_{T_z\pm 1, T_z\pm 1} |{\tilde \psi} \rangle =
% \nonumber \\ &\,&
 C^{TT_z}_{T'T_z\pm 1\, 1\mp 1} \sum_\nu C^{TT_z}_{T'T_z - \nu \, 1 \nu}
\langle \varphi | \hat T_{1\, \nu} \hat P^{T'}_{T_z - \nu , T_z\pm 1} |{\tilde \psi} \rangle .
\end{eqnarray}
For unpaired Slater determinants considered here,
the double integral over the isospace Euler angles in Eq.~(\ref{mate}) can be
further reduced
to a one-dimensional integral over the angle $\beta_T$
using  the identity
\begin{equation}
\hat T_{\lambda\mu} e^{i\alpha \hat T_z} = e^{-i\alpha \mu }  e^{i\alpha \hat T_z} \hat T_{\lambda\mu},
\end{equation}
which is the one-dimensional version of the transformation rule (\ref{trans}) valid for
rotations around the $Oz$ axis in the isospace. The final expression
for the matrix element in Eq.~(\ref{mate1}) reads:
%\begin{widetext}
\begin{eqnarray}\label{integ}
\langle \varphi |
\hat T_{1\, \nu} \hat P^{T'}_{T_z - \nu , T_z\pm 1} |{\tilde \psi} \rangle =
\frac{2T'+1}{2} \int_0^\pi  d\beta_T \, \sin\beta_T\,
d^{T'}_{T_z-1, T_z\pm 1} \langle \varphi |
\hat T_{1\, \nu} e^{-i\beta_T \hat T_y} |{\tilde \psi} \rangle
\nonumber \\
= (-1)^\nu \frac{2T'+1}{2} \int_0^\pi d\beta_T \sin\beta_T d^{T'}_{T_z-1, T_z\pm 1} {\cal N} (\Omega, \beta_T)
\int d^3{\vec r}\, {\tilde{\tilde \rho}}_{1\, -\nu} (\Omega, \beta_T, {\vec r})  ,
\end{eqnarray}
\end{widetext}
where ${\tilde {\tilde \rho}}_{1\nu} (\Omega, \beta_T, {\vec r}) $ is
the isovector transition density, and the double-tilde sign indicates
that the right Slater determinant used to calculate this density
is rotated both in space as well as in isospace:
$|{\tilde {\tilde \psi}} \rangle   =
\hat R(\beta_T )\hat R(\Omega ) | \psi \rangle$. The symbol
${\cal N}(\Omega, \beta_T ) = \langle \varphi | \hat R(\beta_T )\hat R(\Omega ) | \psi \rangle$
denotes the overlap kernel.

Since the natural states have good isospin,
the states (\ref{KTmix}) are free from spurious
isospin mixing. Moreover, since the isospin projection is applied to self-consistent
MF solutions, our model accounts for a subtle balance
between the long-range Coulomb polarization,
which tends to make proton and neutron wave functions different,
and the short-range nuclear attraction, which
acts in an opposite way. The long-range polarization
affects globally all s.p. wave functions.
Direct inclusion of this effect in open-shell heavy nuclei is possible essentially only within
the DFT, which is the only {\it no-core\/} microscopic
framework that can be used there.

Recent experimental
data on the isospin impurity deduced in $^{80}$Zr from the giant dipole
resonance $\gamma$-decay studies~\cite{(Cor11)}  agree well with the
impurities calculated using isospin-projected DFT based on modern Skyrme-force
parametrizations \cite{(Sat09a),(Sat11d)}. This further demonstrates
 that the isospin-projected DFT is capable of capturing the essential piece
of physics associated with the isospin mixing.

\subsection{The choice of Skyrme interaction}

As discussed in Ref.~\cite{(Sat10)}, the isospin projection technique outlined above  does
not yield singularities in energy kernels; hence,
it can be safely executed with all commonly used energy density functionals (EDFs).
However, as demonstrated in Ref.~\cite{(Sat11b)},
the isospin projection alone leads to
unphysically large isospin mixing in odd-odd $N=Z$ nuclei. It has thus been concluded that -- in order to
obtain reasonable results --  isospin projection must be augmented by
 angular-momentum projection. This not only increases the numerical
effort, but also brings back the singularities
in the energy kernels~\cite{(Sat11b)} and thus prevents one from using
the modern parametrizations of the Skyrme EDFs, which all contain
density-dependent terms \cite{(Dob07)}.
Therefore, the only option \cite{(Sat11b)} is to use
the Hamiltonian-driven EDFs. For the Skyrme-type functionals, this leaves
us with one choice: the SV parametrization~\cite{(Bei75)}.
In order to better control the time-odd fields,  the standard SV parametrization must be augmented
by the tensor
terms, which were neglected in the original work~\cite{(Bei75)}.

This density-independent parameterization of the
Skyrme functional has the isoscalar effective mass as low as
$\frac{m^*}{m}\approx 0.38$, which is required to reproduce  the actual nuclear saturation properties. The unusual saturation mechanism of SV has a  dramatic impact on the
overall spectroscopic quality of this force, impairing such key properties like
the symmetry energy~\cite{(Sat11b)}, level density, and level ordering. These
deficiencies  also affect the calculated isospin mixing,
which is a prerequisite for realistic estimates of $\delta_{\rm C}$. In particular, in the case of
$^{80}$Zr discussed above, SV  yields
 $\alpha_{\rm C} \approx
2.8$\%, which is considerably smaller than the mean value of
$\bar\alpha_{\rm C} \approx 4.4\pm 0.3$\% obtained by averaging over nine
popular Skyrme EDFs including the MSk1, SkO', SkP, SLy4, SLy5,
SLy7, SkM$^*$, SkXc, and SIII functionals, see Ref.~\cite{(Sat11d)} for further details.
Even though the ISB corrections $\delta_{\rm C}$ are primarily sensitive
to  {\em differences} between isospin mixing in isobaric analogue
states, the lack of a reasonable Hamiltonian-based
Skyrme EDF is probably the most critical deficiency of the current formalism.

The aim of this study is (i) to provide the most reliable set of the
ISB corrections that can  be obtained within the current angular-momentum and isospin-projected
single-reference DFT, and (ii) explore
the sensitivity
of results to  EDF parameters, choice of particle-hole
configurations, and structure of
time-odd fields that correlate valence neutron-proton pairs in odd-odd $N=Z$ nuclei.
In particular, to quantify uncertainties  related to
the Skyrme coupling constants, we have developed a new density-independent variant of the
Skyrme force dubbed hereafter SHZ2, see Table~\ref{tab1}.

The force was optimized purposefully to properties of light magic
nuclei below $^{100}$Sn. The coupling constants $t_0$,
$t_1$,$t_2$,$x_0$ of SHZ2 were found by means of a $\chi^2$
minimization to  experimental~\cite{(Aud03)} binding energies of five
doubly-magic nuclei: $^{16}$O, $^{40}$Ca, $^{48}$Ca, $^{56}$Ni, and
$^{100}$Sn. The procedure reduced the $\chi^2$ from $\sim 6.0$ for SV
set to $\sim 3.6$ for SHZ2. Most of the nuclear matter
characteristics calculated for both sets are similar.
It appears, however, that the fit to light nuclei only  weakly
constrains the symmetry energy.  The bulk symmetry energy of SHZ2
is $a_{sym} \approx 42.2$\,MeV, i.e., it overestimates the
accepted value $a_{sym} \approx 32\pm 2$\,MeV  by  almost 30\%. While this
property essentially precludes using SHZ2 in detailed nuclear structure
studies, it also creates an interesting opportunity for investigating the
quenching of  ISB effects due to the large isospin-symmetry-restoring
components of the force.

\begin{table}[ht]
    \caption{\label{tab1} Skyrme parameters $t_i, x_i$ ($i = 0,1,2,3$), and $W$ of
             SV \cite{(Bei75)}  (second column) and  SHZ2 (third column).
             The last column shows relative changes of parameters (in percent).
             Both parametrizations use the nucleon-mass parameter of
             $\hbar^2/2m=20.73$\,MeV\,fm$^2$. Parameters not listed are equal to zero.
            }
  \begin{ruledtabular}
    \begin{tabular}{crrr}
        \rule[-0.6em]{0pt}{1.8em}param.
               &        SV    &      SHZ2
                                               &   change (\%)     \\
        \hline\hline
        \rule{0pt}{1.2em}$t_0$
               & $-$1248.290  & $-$1244.98830   &  $-$0.26          \\
        $t_1$  &     970.560  &     970.01156   &  $-$0.06          \\
        $t_2$  &     107.220  &      99.50197   &  $-$7.20          \\
%       $t_3$  &       0.000  &       0.00000   &     $--$          \\
        $x_0$  &    $-$0.170  &       0.01906   &$-$111.21          \\
%       $x_1$  &       0.000  &       0.00000   &     $--$          \\
%       $x_2$  &       0.000  &       0.00000   &     $--$          \\
%       $x_3$  &       0.000  &       0.00000   &     $--$          \\
        $W$    &     150      &     150         &     $0$
    \end{tabular}
 \end{ruledtabular}
\end{table}

\subsection{Numerical details}\label{basis}

All calculations presented below  were
done by using the code HFODD~\cite{(Dob09d),(Sch12)}, version (2.48q) or higher, which
includes both the angular-momentum and isospin projections.
In order
to obtain converged results for isospin mixing with respect to basis
truncation, in our SV calculations we used $N=10$ harmonic oscillator (HO) shells for $A < 40$ nuclei,
12  shells for $40 \le A < 62$ nuclei, and 14
shells for $A \ge 62$ nuclei. In SHZ2 test calculations, we took  $N=10$
shells for $A< 40$ and $N=12$  shells for $A> 40$ nuclei.

For the numerical integration over the Euler angles in space and
isospace ($\alpha, \beta, \gamma ; \beta_T$) we used the
Gauss-Tchebyschev (over $\alpha$ and $\gamma$) and Gauss-Legendre
(over $\beta$ and $\beta_T$) quadratures. We took  $n_\alpha = n_\beta
= n_\gamma =20$ and $n_{\beta_T} = 8$ (or 10) integration points.
This choice guarantees that  the calculated  values of $\delta_{\rm C}$  are  not affected by the numerical integration error.

\section{ISB corrections to the superallowed $\bm{0^+  \rightarrow 0^+}$
Fermi transitions}
\label{sec03}

The $0^+  \rightarrow 0^+$ Fermi $\beta$-decay proceeds between
the ground state (g.s.) of the even-even nucleus
$| I=0, T\approx 1, T_z = \pm 1 \rangle$ and its isospin-analogue partner
in the $N=Z$ odd-odd nucleus, $|I=0, T\approx 1, T_z = 0 \rangle$. The corresponding transition matrix element is:
\begin{equation}\label{fermime}
M_{\rm F}^{(\pm )} =
\langle I=0, T\approx 1,
T_z = \pm 1 | \hat T_{\pm} | I=0, T\approx 1, T_z = 0 \rangle.
\end{equation}

\begin{figure}[htb]
\includegraphics[width=0.8\columnwidth]{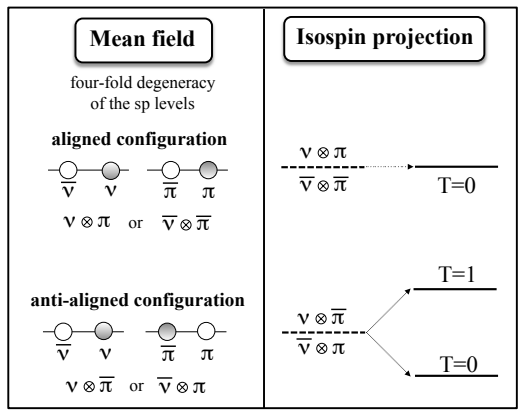}
\caption{Left: two possible g.s.\
configurations of an odd-odd $N$=$Z$ nucleus, as described by the
conventional deformed MF theory. These degenerate configurations are
called aligned (upper) and anti-aligned (lower), depending on what
levels are  occupied by the valence particles. The right panel shows
what happens when the isospin-symmetry is restored. The aligned
configuration is isoscalar; hence, it is insensitive to the isospin
projection. The anti-aligned configuration represents a mixture of
 $T$=0 and $T$=1 states. The isospin projection removes the
degeneracy by lowering the $T$=0 level.
}\label{fig01}
\end{figure}

The  g.s. state $| I=0, T\approx 1, T_z = \pm 1 \rangle$ in Eq.~(\ref{fermime}) is approximated  by a  projected state
\begin{equation}
  | I=0, T\approx 1, T_z = \pm 1 \rangle
  = \sum_{T\geq 1} c^{( \psi )}_{T} \hat P^T_{\pm 1, \pm 1}
     \hat P^{I=0}_{0,0} |\psi \rangle,
\end{equation}
where $|\psi \rangle$ is  the   g.s. of the even-even nucleus obtained in self-consistent MF calculations.
The state $|\psi \rangle$ is unambiguously defined by filling in the pairwise
doubly degenerate levels of protons and neutrons up to the Fermi level.
The daughter state  $|I=0, T\approx 1, T_z = 0 \rangle$ is approximated by
\begin{equation}\label{oddphi}
  | I=0, T\approx 1, T_z = 0 \rangle
  = \sum_{T\geq 0} c^{( \varphi )}_{T} \hat P^T_{0, 0}
     \hat P^{I=0}_{0,0} |\varphi \rangle,
\end{equation}
where  the self-consistent Slater determinant $|\varphi \rangle \equiv |\bar \nu \otimes \pi \rangle$ (or  $| \nu \otimes
\bar \pi \rangle$)
represents the so-called anti-aligned configuration,
selected by placing the odd neutron and the odd proton in
the lowest available time-reversed (or signature-reversed) s.p.\
orbits.
The  s.p.\ configuration
$|\bar \nu \otimes \pi \rangle$ manifestly breaks the
isospin symmetry as schematically depicted
in Fig.~\ref{fig01}. The isospin projection from $|\varphi \rangle$  as expressed  by Eq.~(\ref{oddphi}) is essentially the only way
to reach the $|T\approx 1, I=0\rangle$  states in odd-odd $N=Z$
nuclei.

\subsection{Shape-current orientation}
\label{sec03a}

\begin{figure}[htb]
\includegraphics[width=0.9\columnwidth]{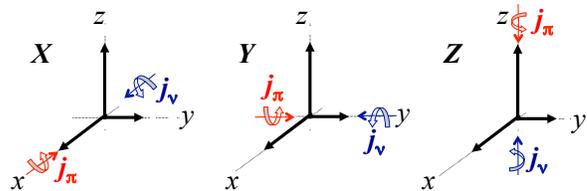}
\caption{(Color online) Schematic illustration of relative orientations of shapes and currents
in the three anti-aligned states $|\varphi_{\rm X}\rangle$ ($X$),
$|\varphi_{\rm Y}\rangle$ ($Y$), and $|\varphi_{\rm Z}\rangle$ ($Z$) discussed in the text.
The long ($Oz$), intermediate ($Ox$), and short
($Oy$) principal axes of the nuclear mass distribution are indicated by thick arrows. The odd-neutron ($j_\nu$) and odd-proton ($j_\pi$) angular momentum
oriented along the $Ox$, $Oy$, or $Oz$ axes is shown by thin arrows. Note that in each case the
total angular-momentum alignment, $j_\nu+j_\pi$, is  zero.
}
\label{fig02}
\end{figure}
\begin{figure}[htb]
\includegraphics[width=0.9\columnwidth]{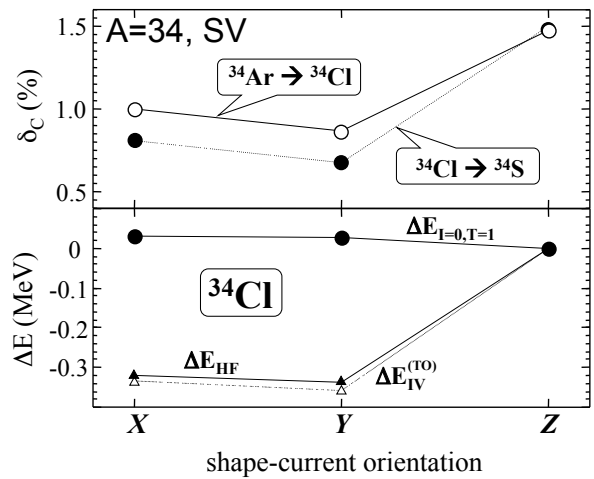}
\caption{Upper panel: the ISB corrections $\delta_{\rm
C}$ for the $0^+ \rightarrow 0^+$ superallowed $\beta$-decays
$^{34}$Ar$\rightarrow ^{34}$Cl (open circles) and
$^{34}$Cl$\rightarrow ^{34}$S (full circles) determined for the
 shape-current orientations  $X$, $Y$, and $Z$ depicted schematically in Fig.~\protect\ref{fig02}.
Lower panel: differences between the energies of the $X$
and $Y$ configurations, and the $Z$ configuration in $^{34}$Cl. Full
triangles correspond to the total HF
energies and open triangles correspond to
contributions from the time-odd isovector
channel. Full dots show the total energy differences obtained for the
angular-momentum and isospin-projected states.
}
\label{fig03}
\end{figure}

%%%
At variance with the even-even parent nuclei, the anti-aligned configurations in odd-odd daughter nuclei
are not uniquely defined.
One of the reasons, which was not fully appreciated in our previous work~\cite{(Sat11c)},
is related to the relative orientation of the nuclear shapes and currents associated
with the valence neutron-proton pairs.
In all signature-symmetry-restricted calculations for triaxial nuclei, such as ours, there are three anti-aligned
Slater determinants with the s.p.\ angular momenta (alignments)
of the valence protons and neutrons pointing, respectively, along
the $Ox$, $Oy$, or $Oz$ axes of the
intrinsic shape defined by means of the long ($Oz$), intermediate ($Ox$), and short
($Oy$) principal axes of the nuclear mass distribution.
These solutions, hereafter referred to as  $|\varphi_{\rm X}\rangle$,
$|\varphi_{\rm Y}\rangle$, and $|\varphi_{\rm Z}\rangle$,  are
schematically illustrated in Fig.~\ref{fig02}. Their properties can be
summarized as follows:
\begin{itemize}

\item
The three solutions are not linearly independent. Their Hartree-Fock
(HF) binding energies may typically differ by a few hundred keV. The differences come
almost entirely from the isovector correlations in the time-odd
channel, as shown in the lower panel of Fig.~\ref{fig03} for a
representative example of $^{34}$Cl. Let us stress that these
poorly-known correlations may significantly impact the ISB
corrections, as shown in the upper panel of Fig.~\ref{fig03}.

\item
The type of the isovector time-odd correlations captured by the HF
solutions depends on the relative orientation of the nucleonic
currents with respect to the nuclear shapes. Solutions oriented
perpendicular to the long axis, $|\varphi_{\rm X}\rangle$ and
$|\varphi_{\rm Y}\rangle$, are usually similar to one another
(they yield identical correlations for axial systems) and
differ from $|\varphi_{\rm Z}\rangle$, oriented parallel to the long axis,
which captures more correlations due to
the current-current time-odd interactions.

\item
The three $|T=1,I=0^+\rangle$ states projected from the $|\varphi_{\rm
X}\rangle$, $|\varphi_{\rm Y}\rangle$, and $|\varphi_{\rm Z}\rangle$
Slater determinants differ in energy by only  a few
tens of keV, see the lower panel of Fig.~\ref{fig03}. Hence,
energy-wise, they represent the same physical solution, differing only slightly due
to the polarization effects originating from different components of the
time-odd isovector fields. However,
since these correlations are completely absent in the even-even
parent nuclei, they strongly impact the calculated
$\delta_{\rm C}$. The largest differences in $\delta_{\rm C}$ have been
obtained  for $A=34$  and $A=74$ systems, see Fig.~\ref{fig03} and
Tables~\ref{tab2} and~\ref{tab3}.

\item
Symmetry-unrestricted calculations always converge to the
signature-symmetry-conserving solution $|\varphi_Z\rangle$ which,
rather surprisingly, appears to be energetically unfavored (except
for $^{18}$F). In spite of our persistent
efforts, no self-consistent tilted-axis solutions have been found.

\end{itemize}

\subsection{Nearly degenerate $\bm{K}$-orbitals}
\label{sec03b}

Owing to an increased density of s.p.\
Nilsson levels in the vicinity of the Fermi surface
for  nearly spherical
nuclei, there appears
another type of ambiguity in choosing the Slater determinants representing
the anti-aligned configurations. Within the set of nuclei studied in this work,
this ambiguity manifests itself particularly strongly in $^{42}$Sc,
where we deal with four possible
anti-aligned MF configurations built on the Nilsson orbits originating
from the spherical $\nu f_{7/2}$ and
$\pi f_{7/2}$ sub-shells. These  configurations can be labeled in terms of the
 quantum number $K$ as $|\nu\bar{K} \otimes \pi K
\rangle $ with $K=1/2$, 3/2, 5/2, and 7/2.

\begin{figure}[htb]
\includegraphics[angle=0,width=0.7\columnwidth]{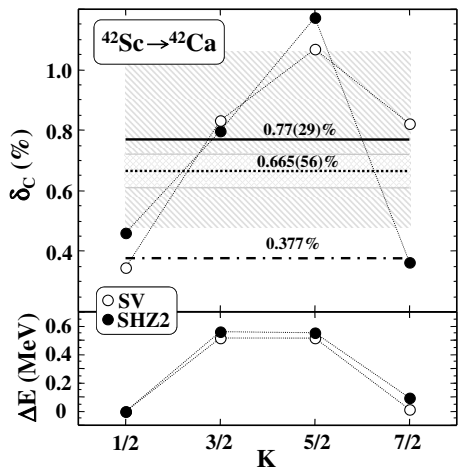}
\caption[T]{\label{fig04}
Top: the ISB corrections to the
$^{42}$Sc$\rightarrow$$^{42}$Ca superallowed $\beta$-transition,
calculated using SV (circles) and SHZ2 (dots) forces by projecting the $|\nu\bar{K} \otimes \pi K \rangle $
configurations in $^{42}$Sc for $K=1/2$, 3/2, 5/2, and 7/2.  From top to bottom, horizontal lines mark
(i) the average ISB correction using SV (thick solid line); (ii) the value of Ref.~\cite{(Tow08)} (dotted line); and (iii)
 $\delta_{\rm C}$ of Ref.~\cite{(Lia09)}. Shaded regions mark the related uncertainties.
Bottom: projected
energies of states $|K; I=0^+, T\approx 1\rangle $ in $^{42}$Sc obtained from the
configurations $|\nu\bar{K} \otimes \pi K \rangle $, relative to the projected energy of the $K=1/2$ state.}
\end{figure}

In the extreme shell-model picture, each of these  states contains
all the $T=1$ and $I$=0, 2, 4, and 6 components. Within the projected
DFT picture, owing to configuration-dependent polarizations in
time-odd and time-even channels, the situation is more complicated
because the Slater determinants $|\nu\bar{K} \otimes \pi K \rangle $
corresponding to different $K$-values are no longer degenerate. Consequently, for each angular momentum $I$, one
obtains four different linearly-dependent solutions.
Calculations show that in all $I$=0 and $T\approx 1$ states of
interest, the isospin mixing $\alpha_{\rm C}$ is essentially
independent of the choice of the initial Slater determinant. In contrast, the calculated ISB corrections
$\delta_{\rm C}$ and energies depend on  $K$, see Fig.~\ref{fig04}.
%%%%

\subsection{Theoretical uncertainties and error analysis}
\label{sec03c}

Based on the discussion presented in Secs.~\ref{sec03a} and
\ref{sec03b}, the recommended calculated values of $\delta_{\rm C}$
for the superallowed $0^+ \rightarrow 0^+$ $\beta$-decay are
determined by averaging over  three relative
orientations of shapes and currents. Only in the case of $A=42$, we
adopt for $\delta_{\rm C}$ an arithmetic mean over the four configurations associated with  different $K$-orbitals.

\begin{figure}[htb]
\includegraphics[angle=0,width=0.7\columnwidth,clip]{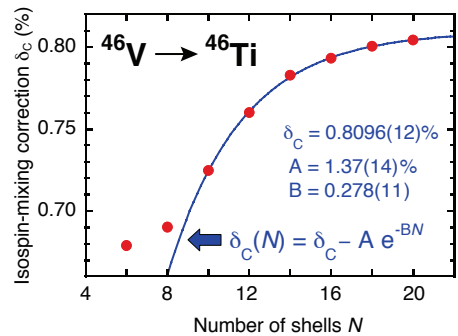}
\caption[T]{\label{fig05}
(Color online) The convergence of the ISB correction $\delta_{\rm C}$
to the $^{46}$V$\rightarrow$$^{46}$Ti superallowed $\beta$-decay
versus the number of oscillator shells considered.}
\end{figure}

To minimize uncertainties in $\alpha_{\rm C}$ and $\delta_{\rm C}$
associated with  the truncation of HO basis in HFODD, we used different HO spaces  in  different mass regions, cf. Sec.~\ref{basis}.
With this choice, the resulting systematic errors due the basis cut-off
should not exceed $\sim 10$\%.
To illustrate the dependence of $\delta_{\rm C}$ on the number of HO shells,
 Fig.~\ref{fig05} shows the case of
the superallowed $^{46}$V$\rightarrow$$^{46}$Ti transition obtained by
projecting from the $|\varphi_{\rm Z}\rangle$ solution in $^{46}$V.
In this case, the parent and daughter nuclei are axial,
which allows us to reduce the angular-momentum projection
to one-dimension and extend the basis size up to $N=20$ HO
shells.

With increasing $N$, $\delta_{\rm C}$ increases, and asymptotically it reaches
the value of $0.8096(12)$\%.  This limiting value is about 6.7\%
larger than the value of $0.7587$\%
obtained for $N=12$ shells, that is, for a basis used to compute
the $42 \leq A\leq 54$ cases.
For $62\leq A\leq 74$ nuclei, which were all found to be triaxial,
we have used $N=14$ shells. The further increase of
basis size  is practically impossible. Nonetheless,
as seen in Fig.~\ref{fig05}, a rate of increase of $\delta_{\rm C}$
slows down exponentially with $N$, which supports
our   10\% error estimate due to the basis truncation.

The total error of the calculated
 value of $\delta_{\rm C}$
includes the standard deviation from the averaging, $\sigma_n$,
and the assumed 10\% uncertainty due to the basis size:
$\Delta (\delta_{\rm C}) = \sqrt{\sigma_n^2  + (0.1\delta_{\rm C})^2 }$.
The same prescription for $\Delta (\delta_{\rm C})$ was also used
in the test calculations with SHZ2,
even though a slightly smaller HO basis was employed in that case.

For  $A=38$ nuclei, our model predicts the   unusually large correction $\delta_{\rm
C}\approx 10$\%.  The origin of a very different isospin mixing
obtained for odd-odd and even-even members of this isobaric triplet  is not fully understood. Most likely, it is a consequence of  the poor spectroscopic properties  of SV. Indeed, as a result of
an incorrect balance between the spin-orbit and tensor terms in SV, the $2s_{1/2}$ subshell  is shifted up in energy close to the Fermi surface.
This state is more sensitive to time-odd polarizations than other
s.p.\ states around $^{40}$Ca core, see Table I in
Ref.~\cite{(Zal08)}.  The calculated equilibrium deformations  $(\beta_2,
\gamma)$  of the $T_z\pm 1$ and $T_z=0$ $A=38$  isobaric  triplet  are very similar, around (0.090,
60$^\circ$). In the following, the
$^{38}$K$\rightarrow$$^{38}$Ar transition is excluded from the
calculation of the $V_{\rm ud}$ matrix element.

\subsection{The survey of ISB corrections in $10\leq A \leq 74$  nuclei}
\label{sec03d}

\begin{table*}
\caption[A]{\label{tab2} Results of calculations for the superallowed transitions measured
experimentally. Shown are: the empirical $ft$-values \cite{(Tow10)}; SV values of $\delta_{\rm C}$
calculated by projecting from the $|\varphi_{\rm
X}\rangle,|\varphi_{\rm Y}\rangle$, and $|\varphi_{\rm Z}\rangle$
Slater determinants,  see Sec.~\protect\ref{sec03a};
recommended mean $\delta_{\rm C}^{{\rm (SV)}}$ corrections
(see Sec.~\protect\ref{sec03c}) and the
corresponding ${\cal F}t$-values; empirical $\delta_{\rm C}^{{\rm (exp)}}$ corrections
calculated by using Eq.~(\protect\ref{expdelt}); contributions
coming from the individual transitions to the $\chi^2$ budget in the
confidence-level test; mean
$\delta_{\rm C}^{{\rm (SHZ2)}}$ corrections and the corresponding
${\cal F}t$-values.}
\begin{ruledtabular}
\begin{tabular}{lccrrrcrrcrrcrr}
Parent    &   $ft$   & $\quad$ &
$\delta_{\rm C}^{\rm (X)}$ & $\delta_{\rm C}^{\rm (Y)}$ &$\delta_{\rm C}^{\rm (Z)}$  & $\quad$ &
$\delta_{\rm C}^{{\rm (SV)}}~\strut$  &  ${\cal F}t~~~~~\strut$  &  $\quad$ & $\delta_{\rm C}^{{\rm (exp)}}~\strut$  &  $\chi^2_i$
& $\quad$  & $\delta_{\rm C}^{{\rm (SHZ2)}}~~\strut$  &  ${\cal F}t~~~~~\strut$ \\
nucleus   &   (s)    &              &
(\%)                   &       (\%)             &     (\%)               &              &
(\%)~~\strut              &      (s)~~~~~\strut      &               &   (\%)~~\strut      &      &
      &       (\%)~~~~\strut        &    (s)~~~~~\strut     \\
\hline
$T_z=-1:$ &              &&      &      &      &&           &       &&       &  &&          &\\
$^{10}$C  &  3041.7(43)  && 0.559& 0.559& 0.823&&   0.65(14)& 3062.1(62)  && 0.37(15) &  3.7 && 0.462(65)&  3067.8(49)\\
$^{14}$O  &  3042.3(11)  && 0.303& 0.303& 0.303&&  0.303(30)& 3072.3(21)  && 0.36(06) &  0.8 && 0.480(48)&  3066.9(24)\\
$^{22}$Mg &  3052.0(70)  && 0.243& 0.243& 0.417&&  0.301(87)& 3080.5(75)  && 0.62(23) &  1.9 && 0.342(49)&  3079.2(72)\\
$^{34}$Ar &  3052.7(82)  && 0.865& 0.997& 1.475&&   1.11(29)&   3056(12)  && 0.63(27) &  3.1 &&  1.08(42)&    3057(15)\\[3pt]
$T_z=0: $ &             &&       &      &      &&      &            &&       &  &&          &\\
$^{26}$Al &  3036.9(09) &&  0.308& 0.308& 0.494&&  0.370(95)& 3070.5(31)  && 0.37(04) &  0.0 && 0.307(62)&  3072.5(23)\\
$^{34}$Cl &  3049.4(11) &&  0.809& 0.679& 1.504&&   1.00(38)&   3060(12)  && 0.65(05) & 48.4 &&  0.83(50)&    3065(15)\\
$^{42}$Sc &  3047.6(12) &&    ---&   ---&   ---&&   0.77(27)& 3069.2(85)  && 0.72(06) &  0.5 &&  0.70(32)&    3071(10)\\
$^{46}$V  &  3049.5(08) &&  0.486& 0.486& 0.759&&   0.58(14)& 3074.6(47)  && 0.71(06) &  4.5 && 0.375(96)&  3080.9(35)\\
$^{50}$Mn &  3048.4(07) &&  0.460& 0.460& 0.740&&   0.55(14)& 3074.1(47)  && 0.67(07) &  3.1 &&  0.39(13)&  3079.2(45)\\
$^{54}$Co &  3050.8(10) &&  0.622& 0.622& 0.671&&  0.638(68)& 3074.0(32)  && 0.75(08) &  2.0 &&  0.51(20)&  3078.0(66)\\
$^{62}$Ga &  3074.1(11) &&  0.925& 0.840& 0.881&&  0.882(95)& 3090.0(42)  && 1.51(09) & 44.0 &&  0.49(11)&  3102.3(45)\\
$^{74}$Rb &  3084.9(77) &&  2.054& 1.995& 1.273&&   1.77(40)&   3073(15)  && 1.86(27) &  0.1 &&  0.90(22)&    3101(11)\\[3pt]
          &             &&       &      &     &&
                       $\overline{{\cal F}t}=$ & 3073.6(12)  &&  $\chi^2 =$ &112.2 && $\overline{{\cal F}t}=$ & 3075.0(12) \\
          &             &&
                 &      &     &&
                               $|V_{\rm ud}|=$ & 0.97397(27) &&$\chi_d^2 =$ & 10.2 &&  $|V_{\rm ud}|=$ &  0.97374(27) \\
          &             &&
                 &      &     &&
                                               & 0.99935(67) &&             &    &&                  &  0.99890(67)
\end{tabular}
 \end{ruledtabular}
\end{table*}

\begin{table}
\caption[A]{\label{tab3}Similar as in Table~\protect\ref{tab2}, except for the unmeasured transitions.}
\begin{ruledtabular}
%\begin{tabular}{lrrrcrrcrrcrr}
\begin{tabular}{lcccrcrc}
Parent    &
$\delta_{\rm C}^{\rm (X)}$ & $\delta_{\rm C}^{\rm (Y)}$ &$\delta_{\rm C}^{\rm (Z)}$  & $\quad$ &
$\delta_{\rm C}^{\rm (SV)}$  & $\quad$  & $\delta_{\rm C}^{\rm (SHZ2)}$  \\
nucleus   &
(\%)                   &       (\%)             &     (\%)               &              &
(\%)                   &       &     (\%)          \\
\hline
$T_z=-1:$ &      &      &      &&           &&           \\
$^{18}$Ne & 2.031& 1.064& 1.142&&   1.41(46)&&    0.72(30)\\
$^{26}$Si & 0.399& 0.399& 0.597&&   0.47(10)&&   0.529(77)\\
$^{30}$S  & 1.731& 1.260& 1.272&&   1.42(26)&&    0.98(21)\\[3pt]
$T_z=0: $ &      &      &      &&           &&            \\
$^{18}$F  & 1.819& 0.956& 0.987&&   1.25(42)&&    0.42(24)\\
$^{22}$Na & 0.255& 0.255& 0.535&&   0.35(14)&&   0.216(86)\\
$^{30}$P  & 1.506& 0.974& 1.009&&   1.16(27)&&    0.60(20)\\
$^{66}$As & 0.956& 0.925& 1.694&&   1.19(38)&&    0.64(12)\\
$^{70}$Br & 1.654& 1.479& 1.429&&   1.52(18)&&    1.10(52)
\end{tabular}
 \end{ruledtabular}
\end{table}

\begin{figure}[htb]
\includegraphics[angle=0,width=0.8\columnwidth]{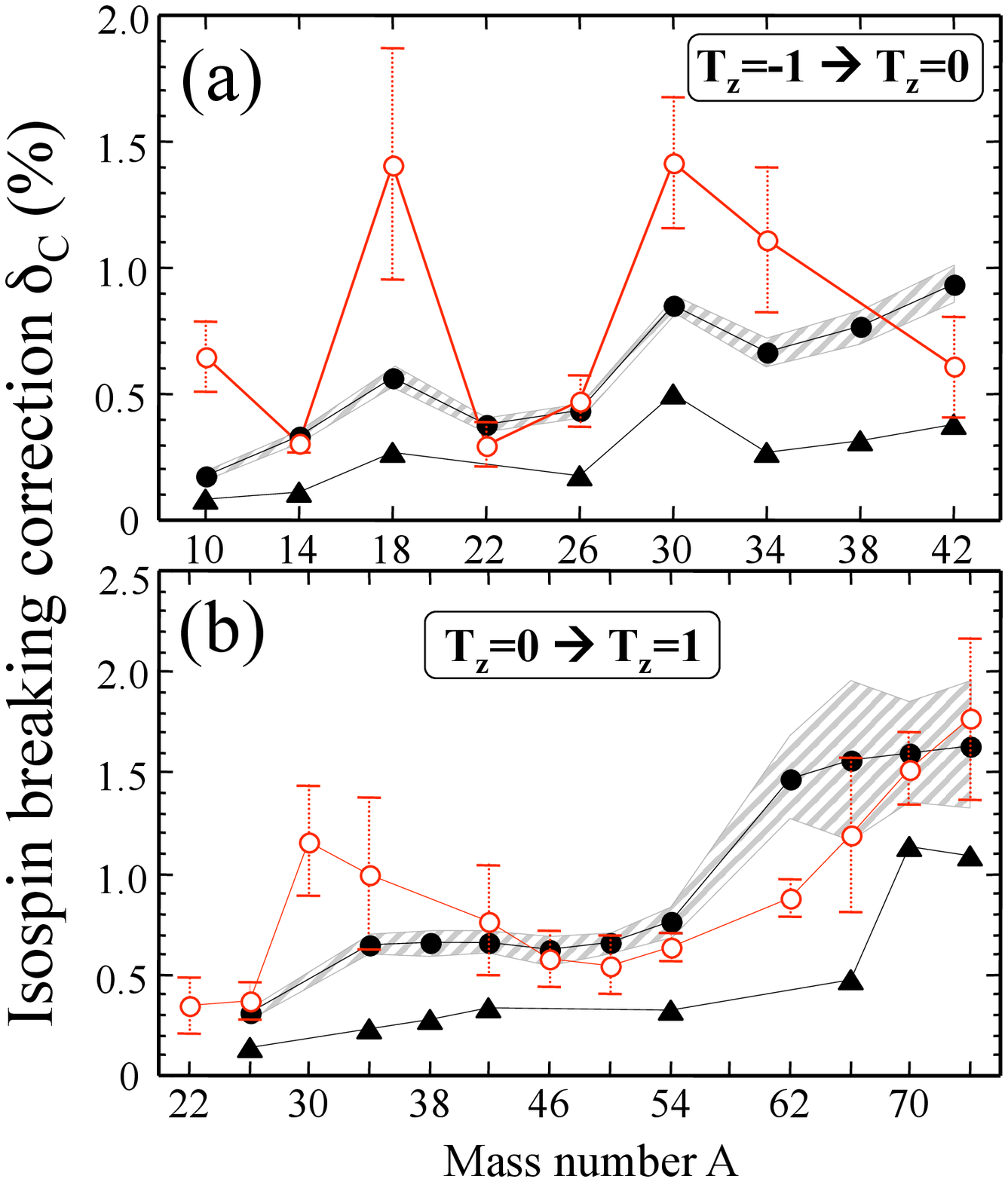}
\caption[T]{\label{fig06}
(Color online) ISB corrections to the superallowed $0^+\rightarrow 0^+$
$\beta$-decays
calculated for (a)  $T_z= -1 \rightarrow T_z =0 $  and (b) $T_z= 0 \rightarrow
T_z =1 $ transitions.
Our adopted values from
 Table~\ref{tab2} (circles with error bars) are compared with
 ISB corrections from Refs.~\cite{(Tow08)} (dots, shaded band marks errors)
and \cite{(Lia09)} (triangles). }
\end{figure}

\begin{figure}[htb]
\includegraphics[angle=0,width=0.7\columnwidth,clip]{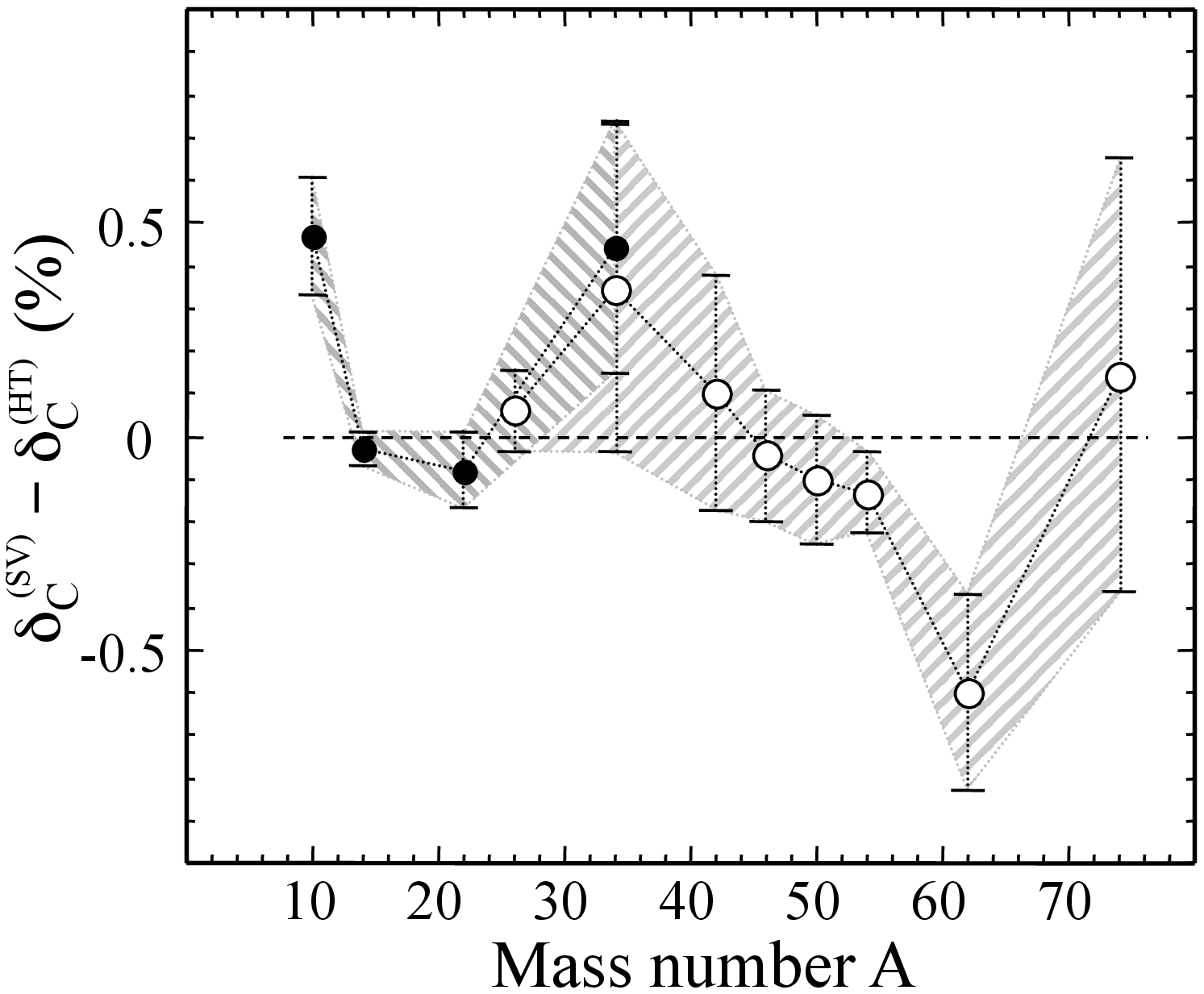}
\caption[T]{\label{fig07}
Differences between the ISB corrections to the twelve accurately
measured superallowed $0^+\rightarrow 0^+$ $\beta$-transitions
(excluding $A$=38) calculated in this work with SV  and those of
HT \cite{(Tow08)}. Circles and dots  mark the $T_z=-1
\rightarrow T_z = 0$ and $T_z= 0 \rightarrow T_z = 1$ decays, respectively.
The errors, calculated as
$\sqrt{(\Delta \delta_{\rm C}^{\rm (SV)})^2 + (\Delta \delta_{\rm
C}^{\rm (HT)})^2}$, are shown by shaded bands.}
\end{figure}

The results of our calculations are collected in Tables~\ref{tab2}
and  \ref{tab3}, and in  Fig.~\ref{fig06}. In addition,
Fig.~\ref{fig07} shows the differences, $\delta_{\rm C}^{({\rm SV})}-
\delta_{\rm C}^{({\rm HT})}$, between our results and those of
Ref.~\cite{(Tow08)}. In spite of clear differences between SV and HT, which can be seen for specific transitions including those for
$A=10$, 34, and 62, both calculations reveal a similar
increase of $\delta_{\rm C}$ versus $A$, at variance with the RPA calculations of
Ref.~\cite{(Lia09)}, which  also yield systematically smaller values.

The ISB corrections used for further calculations
of  $V_{\rm ud}$  are collected in Table~\ref{tab2}.
Let us recall that our preference is to use the
averaged corrections and that  the $^{38}$K$\rightarrow$$^{38}$Ar transition has been disregarded.
All other ingredients needed to compute the ${\cal F}t$-values,
including radiative
corrections $\delta_{\rm R}^\prime$ and $\delta_{\rm NS}$, are  taken
from Ref.~\cite{(Tow08)}, and the empirical $ft$-values are taken from
Ref.~\cite{(Tow10)}. For the sake of completeness, these empirical
$ft$-values are also listed in Table~\ref{tab2}.

\begin{figure}[htb]
\includegraphics[width=0.9\columnwidth]{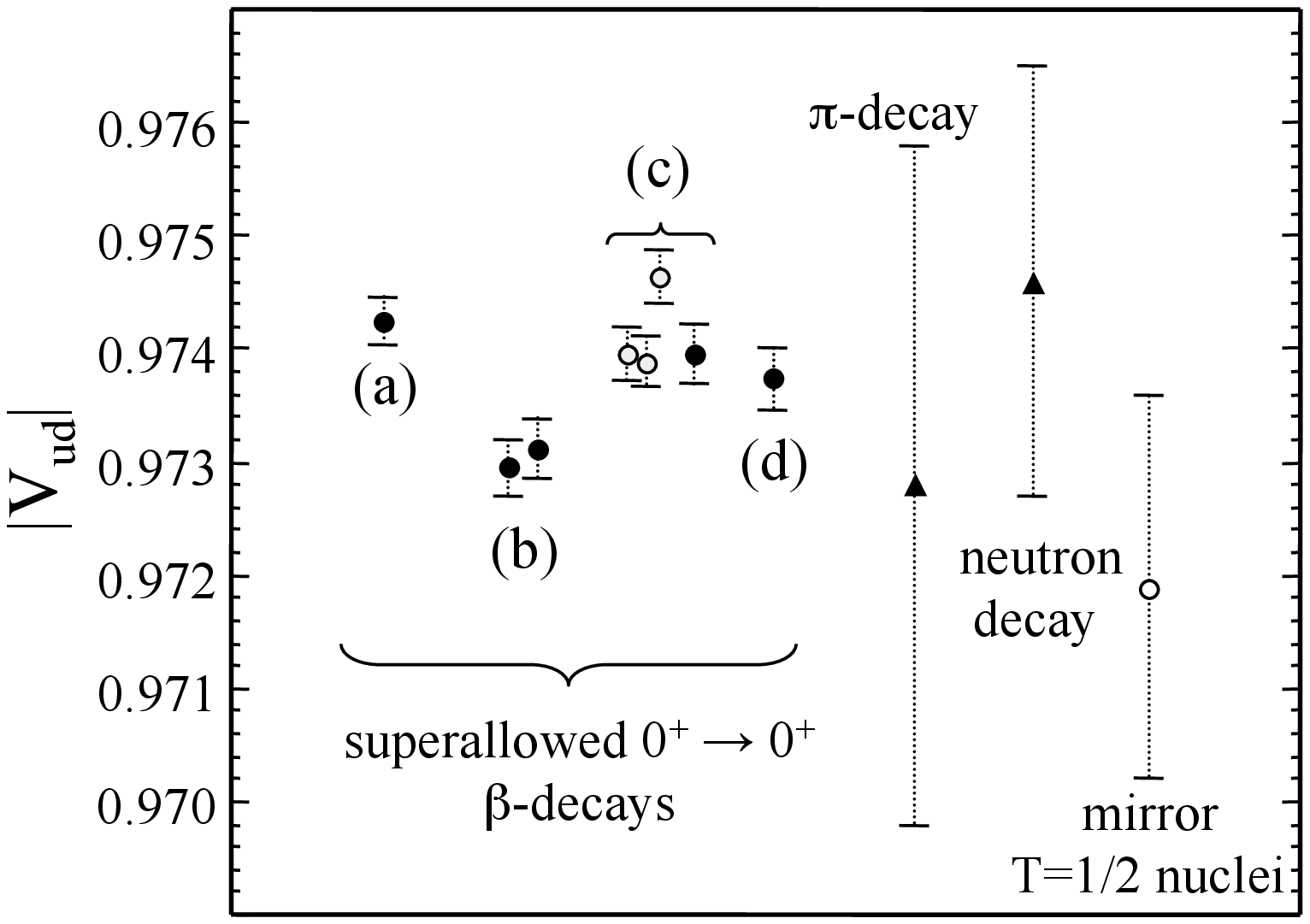}
\caption[T]{\label{fig08}
The matrix element $|V_{\rm ud}|$ deduced from the superallowed
$0^+\rightarrow 0^+$ $\beta$-decay (dots) for different
sets of the $\delta_{\rm C}$ corrections calculated in:
(a) Ref.~\cite{(Tow08)}; (b)  Ref.~\cite{(Lia09)} with NL3
and DD-ME2 Lagrangians; this work, using the averaged values of
$\delta_{\rm C}$ with (c)  SV  and (d) SHZ2
functionals. Gray circles show  ISB
corrections with SV calculated  at fixed
shape-current orientations  $X$, $Y$, and $Z$ (from left to
right). Triangles mark values obtained from
the pion-decay \cite{(Poc04)} and
neutron-decay \cite{(Nak10)} studies. The open
circle shows $|V_{\rm ud}|$  deduced from the $\beta$-decays in the
$T=1/2$ mirror nuclei \cite{(Nav09a)}}.
\end{figure}

\begin{figure}
\includegraphics[angle=0,width=0.9\columnwidth]{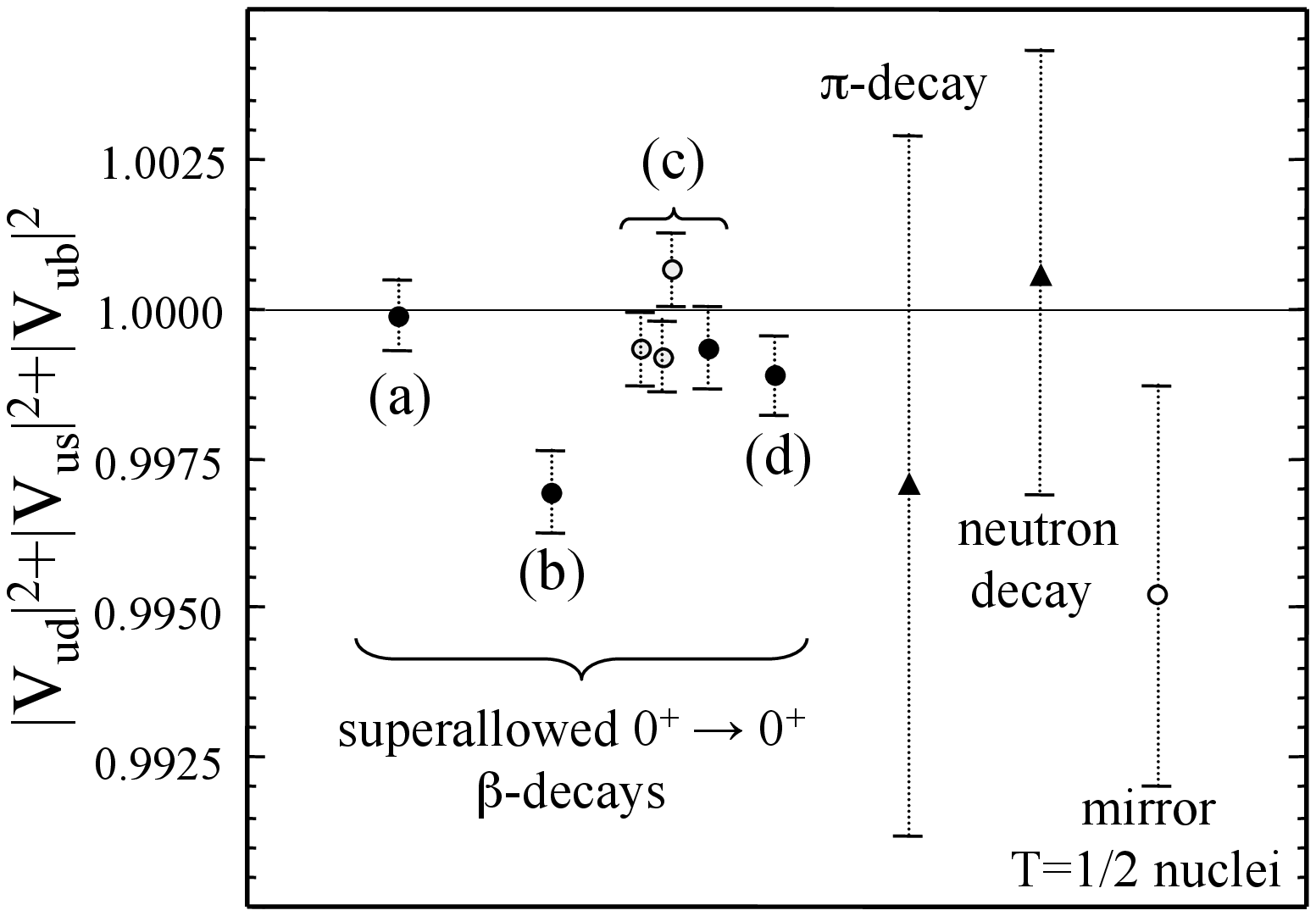}
\caption[T]{\label{fig09}
Similar as   in Fig.~\protect\ref{fig08} except for the unitarity condition
(\ref{ckm}).
}
\end{figure}

In the
error budget  of the
resulting ${\cal F}t$-values  listed in  Table~\ref{tab2},
apart from errors in the $ft$ values
and radiative corrections,  we also included the uncertainties estimated for
the calculated values of $\delta_{\rm C}$, see Sec.~\ref{sec03c}.
To conform with HT, the average value
$\overline{{\cal F}t} = 3073.6(12)$s was calculated by using the
Gaussian-distribution-weighted formula.
However, unlike HT, we do not apply any further corrections to
$\overline{{\cal F}t}$.
This leads to $|V_{\rm ud}| = 0.97397(27)$, which agrees very well
with both the HT result~\cite{(Tow08)},  $|V_{\rm ud}^{{\rm (HT)}}| = 0.97418(26)$,
and the central value obtained from the neutron decay
$|V_{\rm ud}^{(\nu )}| = 0.9746(19)$~\cite{(Nak10)}.
A survey of the $|V_{\rm ud}|$ values deduced by using different  methods is
given in Fig.~\ref{fig08}. By combining the value of $|V_{\rm ud}|$
calculated here with those of $|V_{\rm us}| = 0.2252(9)$ and
$|V_{\rm ub}| = 0.00389(44)$ of the 2010 Particle Data
Group~\cite{(Nak10)}, one obtains
\begin{equation}\label{ckm}
       |V_{\rm ud}|^2 +  |V_{\rm us}|^2  + |V_{\rm ub}|^2 =  0.99935(67),
\end{equation}
which implies that the unitarity of the first row of the  CKM matrix  is satisfied
with a precision better than 0.1\%. A survey of the unitarity condition (\ref{ckm}) is shown in Fig.~\ref{fig09}.

It is worth noting that by using  $\delta_{\rm C}$ values corresponding to the fixed  current-shape
orientations ($|\varphi_{\rm X}\rangle$, $|\varphi_{\rm Y}\rangle$, or
$|\varphi_{\rm Z}\rangle$) instead of their
average, one still obtains compatible results for $|V_{\rm
ud}|$ and unitarity condition (\ref{ckm}), see Figs.~\ref{fig08}
and~\ref{fig09}. Moreover, the value of $|V_{\rm ud}|$ obtained
by using  SHZ2  is only $\approx$0.024\,\% smaller than
the SV result, see Table~\ref{tab2}. This is an intriguing result, which indicates that an increase
of the bulk symmetry energy -- that tends to restore the isospin
symmetry --  is partly compensated by other effects. The most
likely origin of this compensation mechanism is due to the time-odd
spin-isospin mean fields, which  are poorly constrained by the standard
fitting protocols of  Skyrme EDFs \cite{(Ost92w),(Ben02),(Zdu05)}.
For instance, if one compares the Landau-Migdal  parameters characterizing the
spin-isospin time-odd channels \cite{(Ost92w),(Ben02),(Zdu05)} of SV ($g_0 = 0.57$, $g_0^\prime = 0.31$, $g_1 = 0.46$, $g_1^\prime = 0.46$) and SHZ2
($g_0 = 0.27$, $g_0^\prime = 0.30$, $g_1 = 0.47$, $g_1^\prime = 0.47$) one notices that these two functionals   differ by a factor of two in the scalar-isoscalar Landau-Migdal parameter
$g_0$.

\begin{figure}[htb]
\includegraphics[width=0.8\columnwidth]{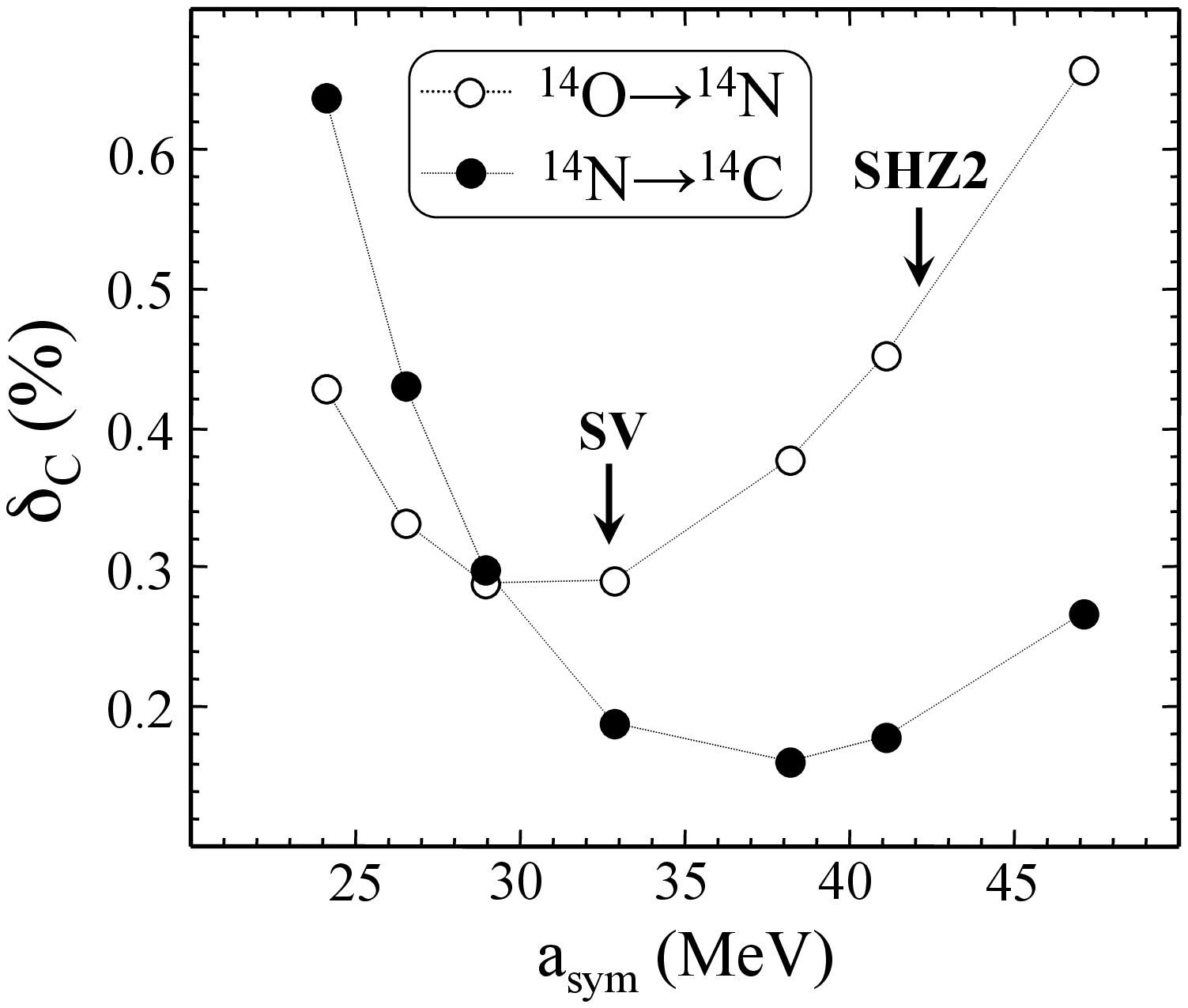}
\caption[T]{\label{fig10}
ISB corrections for $^{14}$O$\rightarrow ^{14}$N and $^{14}$N$\rightarrow ^{14}$C superallowed
$0^+ \rightarrow 0^+$ $\beta$-decays
calculated for a set of SV-based Skyrme forces  with systematically varied $x_0$ parameter,
which affects   the bulk asymmetry energy coefficient $a_{sym}$  and spin-spin fields. The arrows indicate $x_0$ values corresponding
to  SV  and SHZ2.
}
\end{figure}
%%%%
To illustrate the compensation mechanism related to  the bulk symmetry energy
and $g_0$,
 in Fig.~\ref{fig10} we plot  $\delta_{\rm C}$ for the
$^{14}$O$\rightarrow ^{14}$N$ \rightarrow ^{14}$C
super-allowed $0^+ \rightarrow 0^+$ transitions
as  functions of the bulk symmetry parameter $a_{sym}$ for a set of SV-based Skyrme forces  with systematically varied $x_0$ parameter.
At a functional level,  $x_0$  affects only two Skyrme coupling
constants (see e.g. Appendix A in Ref.~\cite{(Ben03)}):
\begin{eqnarray}
C_1^\rho & = & -\frac{1}{4} t_0 \left( \frac{1}{2} + x_0 \right) - \frac{1}{24} t_3 \left(  \frac{1}{2} + x_3 \right)\rho_0^\alpha , \\
C_0^s    & = & -\frac{1}{4} t_0 \left(  \frac{1}{2} - x_0 \right) - \frac{1}{24} t_3 \left(  \frac{1}{2} - x_3 \right) \rho_0^\alpha.
\end{eqnarray}
The  coupling constant $C_1^\rho$ influences the isovector part of the bulk symmetry energy  \cite{(Sat06)}
while $C_0^s$  affects  $g_0$. The ISB correction in Fig.~\ref{fig10}
exhibits a  minimum indicating the presence of the compensation effect.
Similar effect was calculated for the $A=34$ transitions. Hence, it is safe to state
that our exploratory calculations are indicative of  the interplay between the symmetry energy and time-odd fields.

\subsection{Confidence level test}

In this section, we present results of the confidence-level (CL) test
proposed in  Ref.~\cite{(Tow10)}. The CL test
is based on the assumption that the CVC hypothesis is valid
up to at least $\pm 0.03$\%, which implies that a set of
structure-dependent corrections should produce statistically consistent
set of ${\cal F}t$-values. Assuming the validity of the
calculated corrections  $\delta_{\rm NS}$~\cite{(Tow94)},
the empirical ISB
corrections can be defined as:
\begin{equation}\label{expdelt}
\delta_{\rm C}^{{\rm (exp)}}  = 1 + \delta_{\rm NS}
- \frac{\overline{{\cal F}t}}{ft(1+\delta_{\rm R}^\prime)}.
\end{equation}
By the least-square minimization of the appropriate
$\chi^2$, and treating the value of $\overline{{\cal F}t}$ as a
single adjustable parameter, one can attempt to bring the set of
empirical values $\delta_{\rm C}^{{\rm (exp)}}$ as close as
possible to the set of  $\delta_{\rm C}$.
%%%
\begin{figure}[htb]
\includegraphics[width=0.9\columnwidth]{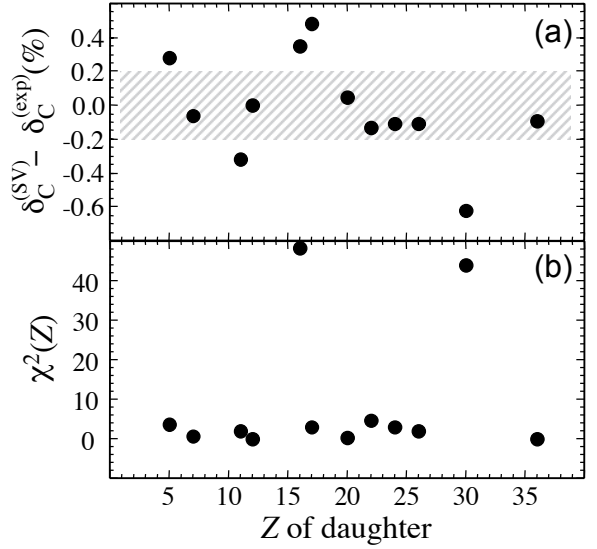}
\caption[T]{\label{fig11}
Top: differences between the calculated
ISB corrections and empirical values resulting from
the CL test of   Ref.~\cite{(Tow10)}. The shaded area of width $\pm 0.2$\%
is added in order to better visualize the differences.
Bottom: contributions from individual transitions to
the  $\chi^2$ budget. Note the particularly large contributions from
the $^{34}$Cl$ \rightarrow ^{34}$S and
$^{62}$Ga $\rightarrow ^{62}$Zn transitions that
deteriorate the CL test. See text for details.
}
\end{figure}
%--------------------------------------------------------------------------

The  empirical ISB corrections
deduced in this way are tabulated in  Table~\ref{tab2} and
illustrated in Fig.~\ref{fig11}.
Table~\ref{tab2} also lists  individual contributions to the $\chi^2$
budget. The obtained $\chi^2$ per degree of freedom ($n_d=11$) is
$\chi^2 /n_d = 10.2$.
This number is twice as large as that quoted in our previous work~\cite{(Sat11c)},
because of the large uncertainty  of $\delta_{\rm C}$
for the $^{34}$Cl$\rightarrow ^{34}$S transition. Other than that, both previous and present
calculations have difficulty in reproducing the strong increase for $A=62$.
Our $\chi^2 /n_d$ is also  higher than the perturbative-model values reported
in Ref.~\cite{(Tow10)}  ($\chi^2 /n_d = 1.5$),
shell model with Woods-Saxon (SM-WS) radial wave functions  (0.4) \cite{(Tow08)},
shell model with Hartree-Fock (SM-HF) radial wave functions  (2.0) \cite{(Orm96),(Har09)}, Skyrme-Hartree-Fock with RPA   (2.1) \cite{(Sag96a)} , and relativistic Hartree-Fock plus RPA
model (RHF-RPA) \cite{(Lia09)}, which yields $\chi^2 /n_d = 1.7$.

It is worth noting that after disregarding the two transitions that strongly violate the CVC hypothesis, $^{34}$Cl$\rightarrow ^{34}$S
and $^{62}$Ga$\rightarrow ^{62}$As that,  and then performing a new CL test for
the remaining ten transitions ($n_d=9$), the normalized $\chi^2$ drops to 1.9. Within this restricted set of data,
the calculated $|V_{ud}| = 0.97420(28)$ and unitarity condition 0.99978(68) almost
perfectly match the results of Ref.~\cite{(Tow08)}.

\subsection{ISB corrections in $78\leq A \leq 98$ nuclei}

\begin{figure}[htb]
\includegraphics[angle=0,width=0.7\columnwidth,clip]{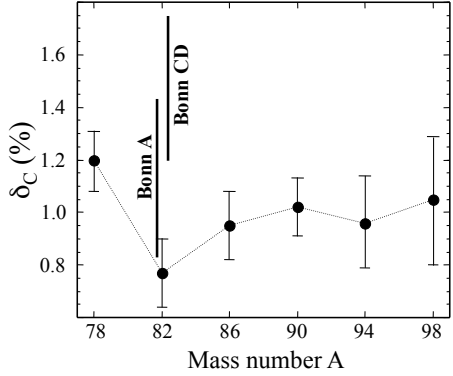}
\caption[T]{\label{fig12}
The
ISB corrections to the superallowed $0^+ \rightarrow 0^+$ transitions
in heavy nuclei calculated in the present work (full dots).
Vertical bars mark the ISB corrections
to the $^{82}$Nb$\rightarrow$$^{82}$Zr
transition calculated in Ref.~\cite{(Pet08)}
by using
the VAMPIR formalism with the charge-independent Bonn A and
charge-dependent Bonn CD interactions.
}
\end{figure}

Our  projected DFT approach can be used to  predict isospin mixing  in heavy nuclei. The
calculated ISB corrections and $Q$-values in $78 \le A \le 98$ nuclei
are listed in Table~\ref{tab4}. The values of $\delta_{\rm C}$ are
also shown in Fig.~\ref{fig12}. Note that the predicted ISB corrections
are here considerably smaller than those in $A$=70
and $A$=74 nuclei, see Tables~\ref{tab2} and \ref{tab3}. For the sake
of comparison,  Fig.~\ref{fig12} also shows
predictions of Ref.~\cite{(Pet08)} for
the $^{82}$Nb$\rightarrow ^{82}$Zr transition  using the VAMPIR approach with either charge-independent Bonn A potential
or charge-dependent Bonn CD potential. Note that our prediction is only slightly below the   Bonn A result
and significantly lower than   the  Bonn CD
value. For the sake of completeness, it should be mentioned that
our $Q_\beta$-value of 10.379\,MeV for this transition agrees well with
 $Q_\beta=10.496$\,MeV (Bonn A) and 10.291\,MeV (Bonn CD) calculated
within the VAMPIR approach.
%%%

\begin{table*}[ht]
\caption[A]{\label{tab4} Results of calculations for the superallowed $\beta$-decays
in $78\leq A \leq 98$ nuclei: the isospin impurities in
the parent and daughter nuclei; $\delta_{\rm C}$  for
different shape-current orientations;  averaged (recommended)  $\delta_{\rm C}$; calculated equilibrium deformations $\beta_2$ and $\gamma$; and  $Q_\beta$-values calculated here and estimated from the
extrapolated masses of Ref.~\cite{(Aud03)}.
}
 \begin{ruledtabular}
\begin{tabular}{ccccccccccccccccc}

           &               &               &$\quad$& {$\alpha_{\rm C}^{{\rm (P)}}$} & {$\alpha_{\rm C}^{{\rm (D)}}$} &$\quad$&
           {$\delta_{\rm C}^{{\rm (X)}}$}      &{$\delta_{\rm C}^{{\rm (Y)}}$}      &{$\delta_{\rm C}^{{\rm (Z)}}$}      &
           {$\delta_{\rm C}^{{\rm (SV)}}$}
           & $\quad$ &  $\beta_2^{{\rm (SV)}}$& $\gamma^{{\rm (SV)}}$
           & $\quad$ &$Q_\beta^{{\rm (th)}}$  & $Q_\beta^{{\rm (exp)}}$ \\
           &               &               &$\quad$&  (\%)   &  (\%)&$\quad$&(\%)  & (\%)   &  (\%)  &   (\%) & $\quad$ &
       & (deg) &$\quad$&(MeV)&  (MeV)           \\
\hline
$T_z= 0 $  & $\rightarrow$ & $ T_z = 1$    &       &         &         & &         &        &        &             & &
       &         & &         &                  \\
$^{78}$Y   & $\rightarrow$ & $^{78}$Sr     &       &  2.765  &  0.976  & &   1.20  &  1.19  &  1.20  &  1.20(12)   & &
0.004  &  60.0   & & 10.471  &  10.650$^{\#}$   \\
$^{82}$Nb  & $\rightarrow$ & $^{82}$Zr     &       &  3.099  &  1.408  & &   0.70  &  0.91  &  0.70  &  0.77(13)   & &
0.036  &  60.0   & & 10.379  &  11.220$^{\#}$   \\
$^{86}$Tc  & $\rightarrow$ & $^{86}$Mo     &       &  3.337  &  1.518  & &   0.89  &  0.89  &  1.08  &  0.95(13)   & &
0.122  &   0.0   & & 10.965  &  11.350$^{\#}$   \\
$^{90}$Rh  & $\rightarrow$ & $^{90}$Ru     &       &  3.525  &  1.608  & &   0.99  &  0.99  &  1.09  &  1.02(11)   & &
0.161  &   0.0   & & 11.465  &  12.090$^{\#}$   \\
$^{94}$Ag  & $\rightarrow$ & $^{94}$Pd     &       &  3.674  &  1.689  & &   0.86  &  0.86  &  1.17  &  0.96(18)   & &
0.136  &   0.0   & & 11.896  &  13.050$^{\#}$  \\
$^{98}$In  & $\rightarrow$ & $^{98}$Cd     &       &  3.805  &  1.771  & &   0.89  &  0.89  &  1.36  &  1.05(25)   & &
0.057  &   0.0   & & 12.343  &  13.730$^{\#}$   \\
\end{tabular}
 \end{ruledtabular}
\end{table*}

Our calculated values of $\delta_{\rm C}$ are in heavy nuclei
considerably smaller than those
obtained from a perturbative expression \cite{(Dam69a),(Tow77),(Tow10)}:
\begin{equation}\label{damg}
\delta_{\rm C} = 0.002645 \frac{Z^2}{A^{2/3}} (n+1)(n+ \ell +3/2)\, {\rm (\%)} ,
\end{equation}
where $n$ and $\ell$ denote the number of radial nodes and angular
momentum of the valence s.p.\ spherical wave function,
respectively. Indeed, assuming the valence $1p_{1/2}$ state in
$A=78$, Eq.~(\ref{damg}) yields $\delta_{\rm C}$=1.54\%. In
heavier nuclei, where the spherical valence state is $0g_{9/2}$, Eq.~(\ref{damg}) gives $\delta_{\rm C}$
 values that increase smoothly from 1.30\% in $A=82$ to 1.64\% in
$A=98$.

\section{ISB corrections to the Fermi matrix elements
in mirror-symmetric $\bm{T=1/2}$ nuclei}
\label{sec05}

Transitions between the isobaric analogue states in mirror nuclei
$|T=1/2, I, T_z= -1/2\rangle$$\rightarrow$$|T=1/2, I, T_z= + 1/2\rangle$ offer an
alternative way to extract the ${\cal F}t$-values \cite{(Sev08)} and
$V_{ud}$  \cite{(Nav09a),(Nav09b)}. Those transitions are mixed Fermi and Gamow-Teller,
meaning that they are mediated by both the vector and axial-vector
currents. Hence, the extraction of $V_{ud}$ requires --
in addition to  lifetimes and $Q$-values --  measuring another observable, such
as the beta-neutrino correlation coefficient, beta-asymmetry, or
neutrino-asymmetry parameter \cite{(Sev06),(Tow10a)}.  Moreover, the method depends on the radiative and ISB
corrections to both the Fermi and Gamow-Teller matrix elements. In
spite of these  difficulties, current precision of
determination of  $V_{ud}$ using the mirror-decay approach  is  similar to
that offered by  neutron-decay
experiments \cite{(Nak10),(Nav09a),(Nav09b)}, see also
Figs.~\ref{fig08} and~\ref{fig09}.

Within our projected-DFT model, we performed systematic calculations of ISB
corrections to the Fermi matrix elements,
$\delta_{\text{C}}^{\text{V}}$, covering the mirror transitions in
all $11\leq A \leq 49$ nuclei. Calculations were based on the Slater
determinants corresponding to the lowest-energy,  unrestricted-symmetry
HF solutions. If the unrestricted-symmetry calculations did not
converge, the projection was applied to the constrained HF solutions
with imposed signature symmetry.  These two types of solutions differ, in
particular, in relative shape-current orientation, which also varies with
$A$ depending on the s.p.\ orbit occupied by an unpaired nucleon.
It should be underlined, however, that the HF solutions corresponding
to the $\beta$-decay partners were always characterized by the same orientation of
the odd-particle alignment with respect to the
body-fixed reference frame.
All calculations discussed in this section were performed by using the full
basis of $N=12$ HO shells and the SV force.

\begin{table*}
\caption[A]{Results of calculations for $|T=1/2, I, T_z= -1/2\rangle$$\rightarrow$$|T=1/2, I, T_z= + 1/2\rangle$    $\beta$-decays between mirror nuclei: theoretical spin and parity assignments;
isospin-mixing coefficients  in the parent and daughter nuclei;
ISB corrections calculated in this work  (asterisks denote results obtained within unrestricted-symmetry calculations); ISB corrections of
Ref.~\cite{(Sev08)};  quadrupole equilibrium deformation
parameters in the parent nuclei; and
theoretical and experimental $Q_\beta$ values.
}
\label{tab5}
\begin{ruledtabular}
\begin{tabular}{cccccccccrrrrrrrr}
   &   &   & $\quad$
           &    $I^\pi$                       & $\quad$ &   {$\alpha_{\rm C}^{{\rm (P)}}$}   & {$\alpha_{\rm C}^{{\rm (D)}}$}   & $\quad$ & {$\delta^{{\text{V}}\, {\rm (SV)}}_{\rm C}$}     & {$\delta_{\rm C}^{{\text{V}}\, {\rm (S)}}$}
           & $\quad$ &  $\beta_2^{{\rm (SV)}}$& $\gamma^{{\rm (SV)}}$
           & $\quad$ &$Q_\beta^{{\rm (th)}}$  & $Q_\beta^{{\rm (exp)}}$ \\
   &   &   & $\quad$
           &                                  & $\quad$ &            (\%)                    &        (\%)
           & $\quad$ &     (\%)               &           (\%)
           & $\quad$ &                        &        (deg)
           & $\quad$ &           (MeV)        &        (MeV)                  \\
\hline
$^{11}$C   & $\rightarrow$ & $^{11}$B      & & $\frac{3}{2}^-$  & &  0.001  &  0.003  & &    0.077 & 0.928 & &
0.320  &  43.8   & &  1.656  &  1.983   \\
$^{13}$N   & $\rightarrow$ & $^{13}$C      & & $\frac{1}{2}^-$  & &  0.008  &  0.001  & &    0.139 & 0.271 & &
0.210  &  59.1   & &  1.888  &  2.221   \\
$^{15}$O   & $\rightarrow$ & $^{15}$N      & & $\frac{1}{2}^-$  & &  0.012  &  0.002  & &    0.127 & 0.181 & &
0.003  &   0.0   & &  2.446  &  2.754   \\
$^{17}$F   & $\rightarrow$ & $^{17}$O      & & $\frac{5}{2}^+$  & &  0.020  &  0.031  & &    0.167 & 0.585 & &
0.014  &   0.0   & &  2.496  &  2.761   \\
           &                 &             & &                  & &  0.019  &  0.029  & &$^*$0.178 & 0.585 & &
0.064  &  60.0   & &  2.499  &          \\
$^{19}$Ne  & $\rightarrow$ & $^{19}$F      & & $\frac{1}{2}^+$  & &  0.036  &  0.034  & &    0.365 & 0.415 & &
0.321  &   0.0   & &  2.928  &  3.239   \\
$^{21}$Na  & $\rightarrow$ & $^{21}$Ne     & & $\frac{3}{2}^+$  & &  0.047  &  0.052  & &    0.307 & 0.348 & &
0.434  &   0.0   & &  3.229  &  3.548   \\
$^{23}$Mg  & $\rightarrow$ & $^{23}$Na     & & $\frac{3}{2}^+$  & &  0.064  &  0.070  & &    0.340 & 0.293 & &
0.434  &   0.0   & &  3.587  &  4.057   \\
$^{25}$Al  & $\rightarrow$ & $^{25}$Mg     & & $\frac{5}{2}^+$  & &  0.073  &  0.058  & &    0.503 & 0.461 & &
0.444  &   1.6   & &  3.683  &  4.277   \\
$^{27}$Si  & $\rightarrow$ & $^{27}$Al     & & $\frac{5}{2}^+$  & &  0.074  &  0.073  & &    0.472 & 0.312 & &
0.343  &  47.7   & &  4.250  &  4.813   \\
$^{29}$P   & $\rightarrow$ & $^{29}$Si     & & $\frac{1}{2}^+$  & &  0.123  &  0.113  & &    0.694 & 0.976 & &
0.332  &  54.4   & &  4.399  &  4.943   \\
$^{31}$S   & $\rightarrow$ & $^{31}$P      & & $\frac{5}{2}^+$  & &  0.163  &  0.164  & &    0.504 & 0.715 & &
0.315  &   0.0   & &  4.855  &  5.396   \\
$^{33}$Cl  & $\rightarrow$ & $^{33}$S      & & $\frac{3}{2}^+$  & &  0.177  &  0.160  & &    0.644 & 0.865 & &
0.258  &  33.5   & &  5.002  &  5.583   \\
$^{35}$Ar  & $\rightarrow$ & $^{35}$Cl     & & $\frac{3}{2}^+$  & &  0.186  &  0.182  & &    0.576 & 0.493 & &
0.209  &  50.4   & &  5.482  &  5.966   \\
$^{37}$K   & $\rightarrow$ & $^{37}$Ar     & & $\frac{3}{2}^+$  & &  0.291  &  0.267  & &    1.425 & 0.734 & &
0.143  &  60.0   & &  5.589  &  6.149   \\
$^{39}$Ca  & $\rightarrow$ & $^{39}$K      & & $\frac{3}{2}^+$  & &  0.318  &  0.289  & &$^*$0.392 & 0.855 & &
0.034  &  60.0   & &  6.084  &  6.531   \\
$^{41}$Sc  & $\rightarrow$ & $^{41}$Ca     & & $\frac{7}{2}^-$  & &  0.341  &  0.345  & &$^*$0.426 & 0.821 & &
0.032  &  60.0   & &  5.968  &  6.496   \\
$^{43}$Ti  & $\rightarrow$ & $^{43}$Sc     & & $\frac{7}{2}^-$  & &  0.376  &  0.380  & &$^*$0.463 & 0.500 & &
0.090  &  60.0   & &  6.225  &  6.868   \\
$^{45}$V   & $\rightarrow$ & $^{45}$Ti     & & $\frac{7}{2}^-$  & &  0.437  &  0.424  & &    0.534 & 0.865 & &
0.233  &   0.0   & &  6.563  &  7.134   \\
           &                 &             & &                  & &  0.438  &  0.427  & &$^*$0.661 & 0.865 & &
0.233  &   0.0   & &  6.559  &          \\
$^{47}$Cr  & $\rightarrow$ & $^{47}$V      & & $\frac{3}{2}^-$  & &  0.480  &  0.457  & &    0.518 &   --- & &
0.276  &   0.0   & &  6.827  &  7.452   \\
           &                 &             & &                  & &  0.483  &  0.463  & &$^*$0.710 &   --- & &
0.275  &   0.0   & &  6.826  &          \\
$^{49}$Mn  & $\rightarrow$ & $^{49}$Cr     & & $\frac{5}{2}^-$  & &  0.515  &  0.497  & &    0.522 &   --- & &
0.284  &   0.9   & &  7.054  &  7.715   \\
           &                 &             & &                  & &  0.518  &  0.499  & &$^*$0.681 &   --- & &
0.284  &   0.0   & &  7.053  &          \\
\end{tabular}
 \end{ruledtabular}
\end{table*}

\begin{figure}[htb]
\includegraphics[width=0.8\columnwidth]{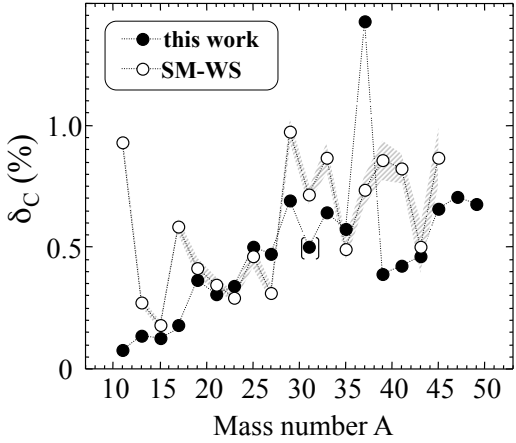}
\caption[T]{\label{fig13}
Full circles: calculated values of the
ISB corrections to the Fermi transitions in $T=1/2$ mirror nuclei.
Open circles with errors: results calculated by
Severijns {\it et al.\/}~\protect{\cite{(Sev08)}}.
}
\end{figure}

The obtained values of the ISB corrections to the
Fermi transitions,
\begin{equation}
  \delta_{\text{C}}^{\text{V}} \equiv 1- | \langle T=\frac{1}{2},I,T_z=\mp \frac{1}{2} | \hat T_\mp | T= \frac{1}{2},I,T_z=\pm \frac{1}{2} \rangle |^2,
\end{equation}
are collected in Table~\ref{tab5} and illustrated in Fig.~\ref{fig13}.
%%%
%%%
Since the calculations were performed in a relatively large basis, the
basis-cut-off-related uncertainty in $\delta_{\text{C}}^{\text{V}}$ could be reduced to
approximately $5\%$, cf.~Sec.~\ref{sec03c}. Except for one case,
theoretical spins and parities of decaying states were taken equal to
those found in experiment:
$I^\pi_{\text{(th)}}=I^\pi_{{\text{(exp)}}}$(g.s.). Only for
$A=31$, no $I=1/2$ component was found in the HF wave
function, and thus the lowest solution corresponding to
$I^\pi_{\text{(th)}}=5/2^+$ was taken instead. It should be mentioned
that, owing to the poor spectroscopic quality of SV,
the projected states corresponding to
$I^\pi_{{\text{(exp)}}}$(g.s.) are not always the lowest ones.
This situation occurs for $A = 19$, 25, and 45, where the lowest states
have  $I^\pi_{\text{(th)}}=5/2^+$, $1/2^+$, and $3/2^-$,
and the corresponding $\delta_{\text{C}}^{\text{V}}$ values are 0.308 \%, 0.419\%, and 0.636\%,
respectively. A relatively strong dependence
of the calculated ISB corrections on spin is worth noting. The calculations also indicate
an appreciable impact of the signature-symmetry constraint on $\delta_{\text{C}}^{\text{V}}$, in particular, in the $pf$-shell nuclei with  $A=45$, 47, and 49. A similar effect was calculated for the $0^+ \rightarrow 0^+$
transitions, see  $\delta_{\rm C}$-values  at fixed
shape-current orientations in Tables~\ref{tab2} and~\ref{tab3}.

\section{The ISB correction to the Fermi decay branch
in $^{32}$C\lowercase{l}}
\label{sec06}

The $V_{\rm ud}$ values extracted by using diverse techniques
including $0^+\rightarrow 0^+$ nuclear decays, nuclear mirror decays,
neutron decay, and  pion decay are subject to both experimental and
theoretical uncertainties. The latter pertain to calculations of
radiative processes and -- for nuclear methods --  to the
 nuclear ISB effect. The uncertainties in radiative and ISB
corrections affect the overall precision of $V_{\rm ud}$
at the level of a few parts per 10$^4$ each \cite{(Har05),(Tow10a)}. It should be stressed,
however, that the ISB contribution to the error bar of $V_{\rm
ud}$ was calculated only for a single theoretical model (SM-WS).
Other microscopic models, including the
SM-HF~\cite{(Har09)}, RH-RPA~\cite{(Lia09)}, and projected
DFT~\cite{(Sat11c)}, yield $\delta_{\rm C}$
corrections that  may  differ substantially from those obtained
in SM-WS calculations.

Inclusion of the model dependence in
the calculated uncertainties is expected to increase the uncertainty
of $V_{\rm ud}$. According to Ref.~\cite{(Lia11)} the increase can
reach even an order of magnitude. In our opinion, a reasonable assessment of
systematic errors (due to the model dependence)  cannot be done at present, as it requires
the assumption that all the nuclear structure models considered are
either equally reliable or their performance can be graded in
an objective way.

A good  way to verify the reliability of various models is to compare
their predictions with empirically determined $\delta_{\rm C}$. Recently,
an anomalously large value of  $\delta_{\rm C}\approx 5.3(9)$\% has been
determined from a precision measurement of the $\gamma$ yields
following the $\beta$-decay of $I=1^+, T=1$ state in $^{32}$Cl to its
isobaric analogue state (Fermi branch) in $^{32}$S~\cite{(Mel11)}.
This value offers a  stringent test on nuclear-structure models,
because it is significantly larger than any value of $\delta_{\rm C}$
in the $A=4n+2$ nuclei. The physical reason for this enhancement
can be traced back to a mixing of two close-lying  $I=1^+$ states
seen in $^{32}$S at the excitation energies of 7002\,keV and 7190\,keV,
respectively~\cite{(Oue11)}. The lower one is the isobaric analogue state
having predominantly $T=1$ component while the higher one is
primarily of $T=0$ character.

\begin{figure}[htb]
\includegraphics[angle=0,width=0.9\columnwidth,clip]{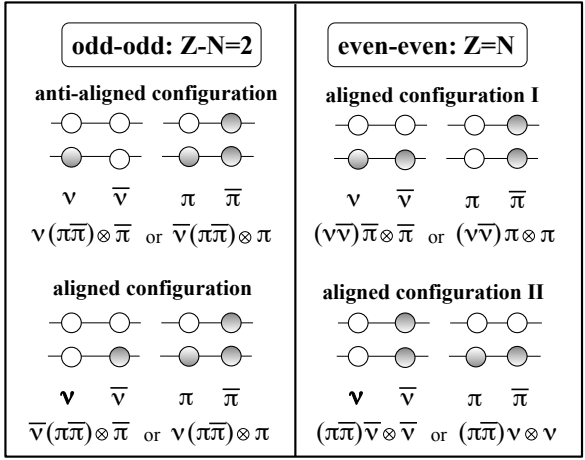}
\caption[T]{\label{fig14}  Schematic illustration of several
possible mean-field configurations in the odd-odd $Z-N=2$ (left) and
even-even $N=Z$ (right) nuclei. The pairs of  proton (neutron)
s.p.\ levels, labeled as $\pi$ and $\bar\pi$ ($\nu$ and $\bar\nu$),
are assumed to be degenerated due to the intrinsic signature symmetry. The orbits $\nu$ and $\pi$
carry the signature quantum number  $r=-i$ $(\alpha =1/2)$ while
$\bar\nu$ and $\bar\pi$
have  $r=i$ $(\alpha =-1/2)$.}
\end{figure}

The experimental value $\delta_{\rm C}\approx 5.3(9)$\% is
consistent with the SM-WS calculations: $\delta_{\rm
C}\approx 4.6(5)$\%. In our projected-DFT approach, we also see
fingerprints of the strong enhancement in  $\delta_{\rm
C}$ value in $^{32}$Cl as compared to  other
$A=4n+2$ nuclei. Unfortunately, a static DFT approach based on
projecting from a single reference state is not sufficient to give a
reliable prediction. This is because, as sketched in Fig.~\ref{fig14}, there exist ambiguities in
selecting the HF reference state. In
the extreme isoscalar s.p.\ scenario, by distributing four valence protons and neutrons over the
  Nilsson s.p.\ levels in an  odd-odd nucleus, one can form two
distinctively different s.p.\ configurations, see Fig.~\ref{fig14}.

The total signature of valence particles determines the total
signature of the odd-odd nucleus and, in turn, an
approximate angular-momentum distribution in its wave
function \cite{(Ben78a)}; the total additive signature $\alpha_{\rm T}\rm{(mod 2)} =
0(1)$ corresponds then to even (odd) spins in the wave
function~\cite{(Boh75)}. It is immediately seen that the
anti-aligned configuration shown in Fig.~\ref{fig14} has $\alpha_{\rm
T} = 0$; hence, in the first approximation, it can be
disregarded. In this sense, the reference wave function in $^{32}$Cl
(or, in general, in any $N-Z=\pm 2$ odd-odd nucleus) corresponds to
the uniquely defined aligned state.
 As seen in  Fig.~\ref{fig14}, this does not hold for  $^{32}$S (or, in general, for  any $N=Z$
even-even nucleus), where one  must consider two possible Slater
determinants having $\alpha_{\rm T} = 1$, obtained by a suitable
proton or neutron particle-hole excitation.

The above discussion  indicates that, contrary to transitions involving
the odd-odd $N=Z$ nuclei studied in Sec.~\ref{sec03}, those
involving even-even $N=Z$ nuclei cannot be directly treated within
the present realization of the model. To this end, the model requires
enhancements including the configuration mixing (multi-reference DFT). Nevertheless, we have carried out
an exploratory study by independently calculating two ISB corrections
for the two configurations discussed above. These calculations
proceeded in the following way:
\begin{itemize}
\item
We select the appropriate
reference configurations which, in the present case,
are: $\nu [4,5,3,3] \pi [5,6,3,3]$ in $^{32}$Cl and
{$\varphi_{{\rm  I}}$:} $\nu [5,5,3,3] \pi [4,6,3,3]$ and
{$\varphi_{{\rm II}}$:} $\nu [4,6,3,3] \pi [5,5,3,3]$
in $^{32}$S. The labels denote the numbers of neutrons and protons
occupying the lowest Nilsson levels in each parity-signature block
$(\pi , r) = (+, +i), (+, -i), (-, +i), (-, -i)$ counting from the bottom of the
HF potential well, as defined in Ref.~\cite{(Dob00c)}.
\item
We determine the lowest
$|I^\pi=1^+, T\approx 1, T_z=-1\rangle$ and
$|\varphi_i; I^\pi=1^+, T\approx 1, T_z=0\rangle$ ($i={\rm I, II}$) states
by projecting onto subspaces of good angular momentum and isospin, and
performing the $K$-mixing and Coulomb rediagonalization as described in
Sec.~\ref{model}.
\item
Finally, we calculate matrix elements of the Fermi operator
$\hat T_\pm$ and extract $\delta_{\rm C}$.
\end{itemize}

The resulting ISB corrections are $\delta_{\rm C}^{(\varphi_{{\rm I}}
)} = 2.40(24)$\% and $\delta_{\rm C}^{(\varphi_{{\rm II}} )} =
4.22(42)$\% for the $\varphi_I $ and $\varphi_{{\rm II}}$
configurations, respectively. As before, we assumed a 10\% error due
to the basis size ($N=10$ spherical HO
shells). Projections from the same configurations cranked in space to
$\langle \hat J_y \rangle$ = 1$\hbar$ (see discussion in
Ref.~\cite{(Zdu07a)}) leaves ISB  corrections almost unaffected:
 $\delta_{\rm C}^{(\varphi_{\rm I} )} = 2.41(24)$\,\% and
$\delta_{\rm C}^{(\varphi_{\rm {II}} )} = 4.30(43)$\,\%. A simple average value would read $\delta_{\rm C} =
3.4(10)$\%, which is indeed strongly enhanced as compared to the
$A=4n+2$ cases. The obtained central value is smaller than both the empirical value
and the SM-WS result. It is worth noting, however,  that within the stated errors our  mean value 3.4(10)\% agrees
with the SM-WS value 4.6(05)\%.
Whether or not the configuration-mixing calculations would provide a significant enhancement
is an entirely open question.

\section{Summary and perspectives}
\label{sec07}

Within the recently-developed unpaired projected-DFT approach, we carried out systematic calculations of isospin mixing effects  and  ISB corrections to the superallowed ${0^+  \rightarrow 0^+}$ Fermi decays in
$10 \le A \le 74$ nuclei and $\beta$-transitions between the isobaric analogue states in mirror ${T=1/2}$  nuclei with
$11 \le A \le 49$. Our predictions are compared with empirical values and with predictions of other theoretical approaches.
Using isospin-breaking corrections computed in our model,  we show that the unitarity of the CKM matrix is satisfied with a precision better than 0.1\%.
We also provide
ISB corrections for  heavier nuclei with $78 \le A \le 98$ nuclei that can guide future experimental and theoretical studies.

We carefully analyze various  model assumptions impacting
theoretical uncertainties of our calculations: basis truncation, definition of the intrinsic state, and configuration selection. To assess the robustness of our results with respect to the choice of interaction, we compared SV results with predictions of the new force SHZ2 that has been specifically developed for this purpose. The comparison of SV and SHZ2 results suggest that ISB corrections are sensitive to the interplay between the bulk symmetry energy and time-odd mean-fields.

While the overall agreement with the empirical values offered by the projected-DFT approach is very encouraging, and the results are fairly robust, there is a lot of room for systematic improvements.
The main disadvantages of our  model in its present formulation include:
(i) lack of pairing correlations; (ii) lack of  ph interaction
(or functional) of good spectroscopic quality;
(iii) the use of a single HF reference state  that cannot accommodate  configuration mixing effects;
(iv) ambiguities in establishing the HF reference state in odd and odd-odd nuclei
caused by different possible orientations of time-odd currents with respect to
total density distribution.
The work on various enhancements of our model, including the inclusion of
$T=0$ and $T=1$ pairing within the
projected Hartree-Fock-Bogoliubov theory, better treatment of configuration mixing using the  multi-reference DFT, and development of the spectroscopic-quality EDF used in projected calculations, is in progress.

\begin{acknowledgments}

This work was supported in part by the Academy of Finland and
University of Jyv\"askyl\"a within the FIDIPRO programme, and by the Office of
Nuclear Physics,  U.S. Department of Energy under Contract Nos.
DE-FG02-96ER40963 (University of Tennessee) and
DE-SC0008499  (NUCLEI SciDAC-3 Collaboration).
We acknowledge the CSC - IT Center for Science Ltd, Finland, for the
allocation of computational resources.
\end{acknowledgments}

%\bibliography{ncnpa}

%

\end{document}